# Fokker-Planck-Boltzmann Model for the Global Electron Velocity Distribution Function Combining Ohmic and Stochastic Heating

## *Uwe Czarnetzki*

Institute for Plasma and Atomic Physics, Faculty of Physics and Astronomy,
Ruhr-University Bochum, 44801 Bochum, Germany

## ABSTRACT

Under low-collisionality conditions the isotropic part of the electron velocity distribution function in a plasma becomes non-local and the electrons can be described by a single global distribution function $f_0(v)$. This is also the regime required for non-local collisonless (stochastic) heating in oscillating and spatially inhomogeneous electric fields. Solution of the Boltzmann equation under these conditions requires usually computationally involving multi-dimensional PIC/MC simulations. The necessity of multi-dimensional simulation arises mainly from the complicated time and space dependence of the collisionless electron heating process. Here it is shown that a time, volume, und solid angle averaged Fokker-Planck operator for the interaction of electrons with an external field can replace the local Ohmic heating operator resulting from a two-term approximation of the Boltzmann equation. This allows consistent treatment of collisional as well as collisonless heating. This operator combined with the dissipative operators for interaction with neutrals, as resulting from the two-term approximation, plus an additional operator for surface losses provide altogether a kinetic description for the determination of the global static and isotropic distribution function of the system. The new operators are relatively easy to integrate in classical local Boltzmann solvers and allow for a fast calculation of the distribution function. As an example, an operator describing non-local collisionless as well as collisional heating in inductively coupled plasmas (ICPs) is derived. The resulting distribution function can then be used to calculate rates and moments in a fluid model of the plasma, similarly to the common practice in the local and highly collisional case. A certain limitation of the above concept is the necessity of using pre-defined field structures. On the other hand, the high computational speed associated with Boltzmann solvers allows quick comparison of models representing various field structures.

## I. INTRODUCTION

Simulation of plasmas has become quite mature in recent years. However, the kinetic simulation of low pressure plasmas, involving two- or three-dimensional oscillating electric field structures and possibly also some background magnetic field is still a challenge. The major choice of simulation technique for these discharges is particle-in-cell (PIC) / Monte-Carlo (MC) simulation although some codes also solve directly the Boltzmann equation [1], [2], [3], [4]. While the basic concept is rather simple and the technique in principle capable of addressing all spatial-temporal aspects of particle kinetics, its detailed implementation for multi-dimensional simulations is still challenging. Further, computational times are often exceptionally long and can range between days and weeks. This limits very much the applicability of such codes to the detailed study of the physics and restricts its use to a small circle of experts.

On the other hand, fluid dynamic codes are much more mature in the sense of providing convenient platforms for relatively easy use by non-specialists. Many commercial products or open source codes are now available. A certain general drawback of fluid dynamic codes is the use of moments, rates, and transport coefficients based on a Maxwellian distribution function of the electrons with the temperature as the only free parameter. Under the non-equilibrium conditions of low-temperature plasmas this is often not very realistic. This is the case particularly when electron-electron collisions are far less frequent than electron-neutral collisions, which is the typical scenario at low ionization degrees.

In order to bring the codes closer to reality, some codes combine the fluid dynamic particle simulation with a local Boltzmann solver for the electrons [5], [6], [7], [8], [9], [10], [11], [12], [13]. This hybrid approach allows combining the speed and convenience of fluid simulations with a realistic kinetic calculation of the electron distribution function and the related moments defining rates and transport coefficients. The Boltzmann solver is usually run prior of the fluid-dynamic simulation and provides moments of the electron distribution function, e.g. the ionization, dissociation or excitation rates, for different values of the reduced electric field $E/N$, where $E$ is the local electric field and $N$ the density of the neutral gas background. The fluid simulation is then started with some initial values of the moments and yields as an output spatially distributed values of the reduced electric field. According to these values new moments are chosen from the table generated by the Boltzmann solver and the simulation is repeated. Usually, fast convergence is achieved and overall the procedure is significantly faster and easier to use than a PIC/MC simulation. Indeed, much effort has been invested in recent years in the development of new and improved Boltzmann solvers [14], [15], [16].

However, this approach is limited to conditions where locality indeed applies, i.e. where the electron mean free path is much shorter than any gradient length of the electric field and naturally also much shorter than the system size. Typically, this is the case for plasmas at pressures higher than about 10 Pa. At significantly lower pressures two new non-local effects become important and the convenient hybrid approach of a fluid-simulation combined with a local Boltzmann solver breaks down.

At mean free paths of the electrons larger than the system size, more precisely at a correspondingly large energy relaxation length, the electron distribution function becomes non-local. This means that the electron distribution function is only a function of the total energy, i.e. the sum of the kinetic energy and the energy in the plasma potential. Then distribution functions measured at different locations overlap perfectly if shifted by the local plasma potential [17], [18], [19], [20]. In contrast to the high pressure case, where a multitude of locally different distribution functions exists, there is only one global distribution function in the low pressure case. The particular form of this distribution function is determined by the overall energy gain out of the electric field and energy losses by collisions and particle losses to the confining walls. This global aspect seems to simplify conditions since the integrated losses could still in principle be calculated by a local Boltzmann solver. However, the challenge is now in properly describing the energy gain out of the electric field. This energy gain is no longer local and Ohmic, i.e. based on momentum changing collisions with the background gas, but becomes also non-local or, as it is usually termed, stochastic.

Stochastic electron heating in plasmas is based on the combination of spatially inhomogeneous and temporally oscillating high frequency fields, typically in the radio-frequency (RF) range, with the thermal motion of the electrons along these gradients [21], [22], [23], [24], [25], [26], [27], [28], [29], [30]. The spatial structure of the field can lead to substantial non-local energy gain well beyond the local quiver motion in the oscillating field. Interaction of the electrons with the field is usually limited to a small part of the entire plasma. This is often a thin layer close to the wall or more precisely close to a dielectric window through which the electromagnetic field enters the plasma. The interaction range is defined by the penetration depth of the field, e.g. the sheath thickness for capacitively coupled RF plasmas (CCP) or the skin depth for inductively coupled plasmas (ICP). In electron cyclotron resonance heating (ECR), a thin resonance region is usually located within the plasma with the extension defined by the gradient length of an inhomogeneous external magnetic field. In all cases electrons enter and leave the heating region due to their thermal motion and gain on average energy during the transition. Energy gain by stochastic heating is usually non-isotropic, typically in one or two spatial dimensions, depending on the particular field structure. Isotropisation takes place subsequently in the volume by elastic collisions with neutrals. At very high plasma densities also collisions with ions can contribute and additionally electron-electron collisions can lead to thermalisation of the distribution function. Naturally, neither fluid simulations nor a Boltzmann solver including only a local Ohmic heating operator can account for these non-local kinetic heating effects. On the other hand resolving the temporal and spatial structure in PIC/MC simulations is computationally very expensive and results consequently in inconvenient long computation times and often highly specialized codes.

The alternative concept proposed here is in the introduction of a new heating operator for standard Boltzmann solvers that accounts for both, local collisional as well as non-local stochastic heating. This operator together with the volume averaged collisional loss operators, as they are standard in any Boltzmann solver, and a new surface loss operator should then allow fast calculation of the global, static, and isotropic electron distribution function. This global distribution function could then serve as an input for fluid models by calculating rate and transport coefficients. An example which uses an experimentally determined global distribution function to calculate spatially resolved plasma parameters is given in [19].

The goal is therefore in deriving a proper operator for the non-local interaction with the field. The operator must describe properly the spatially and temporally integrated but energetically resolved interaction with the structured field and should include also the action of momentum changing collisions, which can be expected to have some residual contribution. In the following it will be demonstrated that the Fokker-Planck operator (FPO) is well suited for this purpose under very general conditions. The operator is averaged over the entire plasma volume, over time and phase, and also over the full solid angle of the velocity. Therefore, it describes the integrated energy exchange of the global, static, and isotropic electron distribution function.

In the course of the derivation of the operator, it will be shown that Ohmic and stochastic heating follow as limiting cases of the same general description of energy exchange between the field and the electrons. Naturally, the global picture is essential for including non-local heating effects. In the collisional Ohmic limit a global average of the local contributions results. The general concept developed in this work leads to universal integral expressions for arbitrary electric field structures. Therefore, the work required in applying the concept to any particular case is reduced to determining the Fourier transform of the associated field structure and carrying out a certain general integral where the absolute square of the Fourier transform is an input function.

The main limitation of the above concept is in the necessity of using a prescribed electric field distribution. Therefore, the spatial field structure is not calculated self-consistently but relies on some model pictures. On the other hand, computation with a local Boltzmann solver, even under the extended conditions proposed here, is of the order of seconds. This allows fast comparison between different models using field structures of different complexity. In particular the effect of more complex field structures on the global distribution function can be studied. Last not least, comparison with more elaborated PIC/MC simulations or experimental data from e.g. Langmuir probe or Thomson scattering measurements are possible.

The remaining part of the paper is organized as follows: Firstly, the classical Boltzmann equation and the two-term approximation are introduced in section 2.1. This includes the standard separation of the equation into a number of individual operators reflecting certain physical processes. In particular the form of the resulting local Ohmic heating operator is discussed and the limitations of this approach are highlighted. Subsequently, the Fokker-Planck operator is introduced in section 2.2 and the concept of applying it to the interaction of electrons with an electric field is outlined. Further, the overall concept of using a global, i.e. volume averaged, kinetic equation is discussed. The equivalence of the energy moment of the Fokker-Planck operator to a kinetic power balance concept already established in the literature for the collisionless case is demonstrated in section 2.3.

The main part of the paper is devoted to the derivation of the Fokker-Planck operator for the interaction between electrons and a general electric field (section 3). This is carried out in two steps: Firstly, Fourier transform in space and time is used to derive a volume averaged power function, here called *h*-function (section 3.1). The limiting cases of pure stochastic and pure Ohmic heating are derived under general conditions (section 3.2). As a particular example the case of a classical inductive discharge (ICP) with an exponentially decaying field is discussed (section 3.3). It is demonstrated that what is otherwise a lengthy calculation becomes an easy operation using the general results from the previous sections. Secondly, the *h*-function and the derivatives of the Fokker-Planck operator are averaged over the full solid angle (section 3.4). Finally, for the collisional limit the differences between the Fokker-Planck operator and the classical two-term approximation operator are discussed (section 3.5).

Derivation of the surface loss operator is performed in one brief section (section 4). This operator accounts for particle and associated energy loss to the confining surfaces of the plasma. The role of the plasma potential and an iterative concept for its self-consistent determination are discussed. The final conclusions are summarized in section 5. More detailed calculations are moved from the main text to two appendices in order to enhance readability. While Appendix A gives an alternative derivation of the two-term approximation, which leads to a form close to the Fokker-Planck operator, the rather lengthy calculations related to the solid angle average are presented in Appendix B.

## II. BOLTZMANN EQUATION AND FOKKER-PLANCK OPERATOR

### A. Classical Boltzmann two-term description

The interaction of electrons with an electromagnetic field and other particles in the plasma is governed by the Boltzmann equation for the electron velocity distribution function $f(\vec{v})$ (*e,m*: electron charge and mass, $t,\vec{r},\vec{v}$ time, space and velocity coordinates, indices indicate respective derivatives, $\vec{E},\vec{B}$: electric and magnetic field) [31] :

$$\frac{\partial f}{\partial t} + \vec{v} \cdot \nabla_r f - \frac{e}{m}\left(\vec{E} + \vec{v} \times \vec{B}\right) \cdot \nabla_v f = \frac{\partial f}{\partial t}\bigg|_{col} . \tag{1}$$

While the left hand side describes transport and the interaction with the electromagnetic field, interactions with other particles are subsumed symbolically under the collision operator on the right, which is a shorthand notation for the Boltzmann collision integral. In this work only electric fields are considered explicitly but generalization to situations including also static magnetic fields, e.g. in connection with electron cyclotron resonance (ECR) heating, is straight forward. Effects related to the induced magnetic field, like the ponderomotive force, are of second order and can be neglected in first order [32]. Extending the concept to second order is beyond the scope of this work.

The standard approach of solving the local Boltzmann equation for electrons interacting with a homogeneous harmonic electric field $\vec{E} = E_0 \cos\left(\omega_0 t\right)\vec{e}_z$ is by the so called two-term approximation [33], [23]. In this approximation the velocity distribution $f(\vec{v})$ is expanded into an infinite series of Legendre polynomials $P_i\left(\cos\left(\vartheta\right)\right)$, where $\cos\left(\vartheta\right) = v_z / v$ and $\vartheta$ is the angle to the axis defined by the direction of the electric field. All coefficients in the expansion are functions of the absolute velocity $v$ only. Under conditions of weak anisotropy the series is terminated after the first order term, so that the expansion contains only two terms, which coins the name: $f\left(\vec{v}\right) \approx f_0\left(v\right) + \cos\left(\vartheta\right)f_1\left(v\right)$. The first term describes the isotropic part and the second term the anisotropic part of the distribution. Together with the Boltzmann collision integral this yields then an equation for the isotropic part of the static distribution function $f_0\left(v\right)$:

$$\left.\frac{\partial f_0}{\partial t}\right|_{TTAO} + \left.\frac{\partial f_0}{\partial t}\right|_{diss} = 0, \tag{2}$$

$$\left.\frac{\partial f_0}{\partial t}\right|_{TTAO} = \frac{v_E^2\,\omega_0}{6\,v^2} \cdot \frac{\partial}{\partial v}\left(g_O\left(\frac{v_m}{\omega_0}\right)v^2 \,\frac{\partial f_0}{\partial v}\right) \tag{3}$$

$$g_O\left(\beta\right) = \frac{\beta}{1+\beta^2}. \tag{4}$$

Here $v_E = e\,E_0 / \left(m\,\omega_0\right)$ is the velocity amplitude of a free oscillating electron. The first term describes the energy input by Ohmic heating and is called the two-term approximation heating operator, indexed as "TTAO". The coupling of energy from the field to the electrons is provided by momentum change in collisions with neutrals with a collision frequency $v_m$. This basic process is represented by the dimensionless Ohmic coupling function $g_O\left(v_m / \omega_0\right)$. The second term, indexed as "diss", results from the Boltzmann collision integral and represents all energy dissipative collision processes, i.e. energy transfer in elastic collisions and inelastic excitation and ionization collisions. Coulomb collisions might be added if necessary, i.e. at elevated electron densities typically in excess of $10^{17}\,m^{-3}$. This is usually done by a Fokker-Planck operator that accounts for stochastic interaction with multiple charges within the Debye sphere [31], [34], [15]. At high charged particle densities the main effect is typically a Maxwellization of the distribution function by electron-electron collisions. Occasionally a Fokker-Planck operator is also introduced in order to describe collisions with atoms [35]. Since inelastic collisions involving discrete states of atoms and molecules are discontinuous and collision operators can contain integrals over the distribution function like in ionization or in Coulomb collisions, special numerical schemes are required for the solution of Eq. (2), e.g. energy grid based relaxation schemes. Instead of solving for the isotropic velocity distribution function $f_0\left(v\right)$, the equation can be rewritten in order to be solved for the energy distribution function $F\left(\varepsilon\right) \sim \sqrt{\varepsilon}\,f_0\left(\sqrt{2\,\varepsilon / m}\right)$, where $\varepsilon = m\,v^2 / 2$ denotes the kinetic energy.

An essential step made by the two-term approximation in transforming the original Boltzmann Eq. (1) to (2) is the separation of the equation into operators with clear functions with respect to energy gain and loss. There is now a heating operator responsible for the energy input and there are energy dissipative operators, according to the various loss channels. The heating operator is purely Ohmic, which means that the coupling between the field and the electrons is local and collisional. The major aim in this work is the replacement of this operator by an alternative one which allows both, Ohmic and non-local stochastic heating.

It is illustrative to evaluate the above Ohmic heating operator (Eq. (3)) for the particular case of a Maxwell-Boltzmann distribution $f_0 \sim \exp\left(-\varepsilon/\left(k_B T_e\right)\right)$ and under the assumption of a constant collision frequency $\nu_m$. $k_B$ is the Boltzmann constant and $T_e$ the electron temperature. This is done in Fig. 1 for the energy distribution function $F(\varepsilon)$ rather than the velocity distribution function $f_0(v)$ since it is more insightful for the present purpose. Naturally, integration over all energies, i.e. the total area under the curve, yields exactly zero, confirming particle number conservation. The zero crossing is at the mean energy $\bar{\varepsilon} = 3/2\, k_B T_e$. Apparently the operator transfers low energy electrons towards higher energies as is expected for a heating process. This general behavior will also be found below for the stochastic heating.

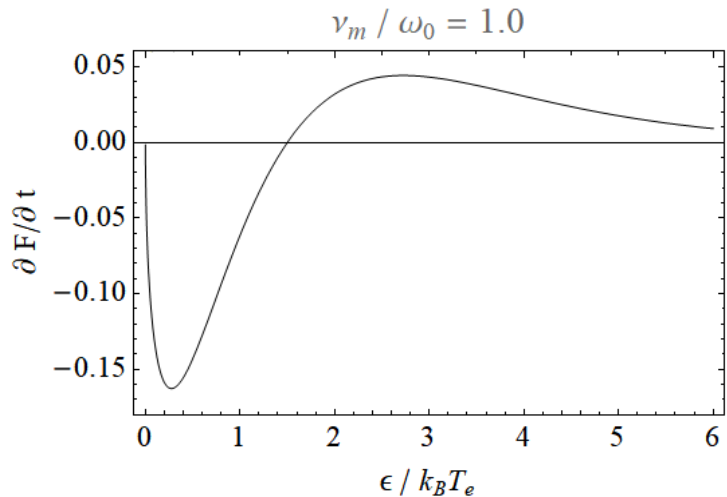

FIG. 1. Ohmic heating operator from the local two-term approximation of the Boltzmann equation acting on a Maxwell-Boltzmann distribution at an electron temperature $T_e$ and for a constant collision frequency $\nu_m$ as a function of the kinetic energy of the electrons $\varepsilon$. The action of the operator is displayed for the energy distribution function $F(\varepsilon)$ and normalized by $m v_E^2\, \omega_0\, n_e/\sqrt{\pi}$, where $n_e$ is the electron density.

An alternative derivation of Eq. (2) is briefly discussed in the following and in more detail in Appendix A. It is interesting that in this alternative approach as an intermediate result, before averaging over the solid angle, an expression very similar to the Fokker-Planck operator results for the case of a homogeneous field. The main idea is that already for an electric field consisting only of a single discrete frequency $\omega_0$ still the distribution function has to be expressed as an infinite series of harmonics of this frequency, i.e. $f = \sum_j f^{(j)} e^{i j \omega_0 t}$, where $j$ is an integer. The cause is in the non-linearity introduced by the term on the lhs of Eq. (1) that contains the product between the electric field and the velocity derivative of the distribution function. The main assumption in the derivation is $v_E \ll v_{th}$, where $v_{th}$ is the characteristic thermal velocity of the distribution, e.g. as following from the mean energy. This ensures fast convergence of the frequency series and allows neglect of higher order terms. The resulting equation for the time independent distribution function $f^{(0)}$ reads:

$$\frac{v_E^2 \, \omega_0}{2} \frac{\partial}{\partial v_z} \left( g_O \left( \frac{v_m}{\omega_0} \right) \frac{\partial f^{(0)}}{\partial v_z} \right) + \frac{\partial f^{(0)}}{\partial t} \bigg|_{diss} = 0, \tag{5}$$

When this equation is averaged over the solid angle, the resulting equation for the isotropic part of the static distribution function is exactly identical to Eq. (3). In the above form the heating term differs from the Fokker-Planck operator, evaluated for the same conditions of a homogeneous field, only by the position of the second derivative with respect to $v_z$. In fact, Eq. (5) is identical to the Fokker-Planck equation for a constant, i.e. velocity independent, collision frequency. The difference is discussed in more detail below in connection with the Fokker-Planck operator.

## B. Fokker-Planck operator

The alternative approach proposed here is based on the Fokker-Planck operator. In plasma physics the Fokker-Planck operator is usually thought to be connected only with Coulomb collisions. Although this is certainly the main application, the concept allows for general stochastic and short time interactions of Markov type [36], [37], [38], [39]. In these interactions the velocity of a particle is changed by an amount $\Delta \vec{v}(\vec{v}, \chi)$. The time interval of the interaction has to be short compared to the time between interactions $\tau = 1/\nu$. Both, $\Delta \vec{v}$ and $\nu$ can depend on the initial velocity $\vec{v}$ and some independent stochastic parameters $\chi = \chi_1, \chi_2, \dots$. In collisions this is typically the impact parameter. By taking the average over these stochastic parameters with the corresponding probability distributions $P(\chi) = P_1(\chi_1) P_2(\chi_2) \dots$, denoted by $\langle \ \rangle_\chi$, the temporal change of the velocity distribution function $f(\vec{v})$ due to the interaction is given by a particular form of the Chapman-Kolmogorov equation. The Fokker-Planck operator follows by expanding the expression to second order in the velocity change $\Delta \vec{v}$, assuming that the distribution function and the collision frequency are only slowly varying functions of the velocity change so that fast convergence is ensured. It can be shown that under certain but quite general conditions higher order terms in the expansion vanish exactly [39], [36]. The resulting Fokker-Planck equation reads:

$$\left. \frac{\partial f}{\partial t} \right|_{FPO} = -\left\langle \frac{\partial}{\partial v_i} \left( \Delta v_i \, \nu(\vec{v}) f(\vec{v}) \right) \right\rangle_\chi + \frac{1}{2} \left\langle \frac{\partial^2}{\partial v_i \partial v_j} \left( \Delta v_i \, \Delta v_j \, \nu(\vec{v}) f(\vec{v}) \right) \right\rangle_\chi . \qquad (6)$$

Here the convention is made that the sum is taken over all Cartesian components $i, j = x, y, z$. The first part describes a drift and the second part diffusion in velocity space. Only this latter part is of relevance in the following since no net-flow or drift is introduced by the oscillating fields in the present context. The operator naturally conserves the particle number for arbitrary normalizeable distribution functions since the integral over velocity space $d^3v$ is zero.

Here, the major aim of applying the Fokker-Planck-Boltzmann concept is the calculation of the non-local isotropic distribution function of the electrons in a low-pressure plasma. The term "low-pressure plasma" is defined as a system where the energy relaxation length of the electrons exceeds the size of the system. The major step is in the derivation of a Fokker-Planck operator for the interaction of the electrons with the electric field and with neutral particles that combines local Ohmic heating, based on collisions, and non-local stochastic heating, based on spatial electric field structures. The Fokker-Planck operator replaces the heating operator derived from the two-term approximation for local Ohmic heating. The global, volume averaged equation to be solved by a local Boltzmann solver then reads:

$$\left. \frac{\partial f_0}{\partial t} \right|_{FFPO} + \left. \frac{\partial f_0}{\partial t} \right|_{dis} + \left. \frac{\partial f_0}{\partial t} \right|_{sur} = 0. \qquad (7)$$

In order to highlight that in the present case the Fokker-Planck operator describes the interaction with the field, the operator is indexed by the acronym "FFPO" for "Field Fokker-Planck Operator". It should be noted that there is some similarity in the approach to the kinetic treatment of wave-particle interactions like in RF-current drive in magnetized fusion devices [40], [41] [42]. However, in that case the field interacts with ions and causes mainly a drift. Further, Ohmic heating is not of relevance and conditions are generally very different.

Further, a new operator $\left. \partial / \partial t \right|_{sur}$ is added in order to account for surface losses, which are not present in a purely local approximation. This operator will be discussed in detail below in section 4. The global and isotropic description is realized by averaging all operators over the entire discharge volume $V$ and further over the full solid angle $\Omega$ associated with the velocity vector, although not marked explicitly here. The dissipative collision operators are the same operators as derived from the two-term approximation. The volume average does not change their local character since the distribution function does not change significantly over the volume. Further, they are isotropic anyway. This allows the use of standard local Boltzmann solvers for the calculation of the global distribution function even in the non-local regime.

Once the new operators are known, the necessary modifications can be considered as minor. The new Fokker-Planck heating operator still remains to be of second order in the derivative. Further, the additional surface loss operator is not much different from an inelastic collision operator. Nevertheless, the main modification is probably introduced by this operator since it contains the plasma potential. This potential needs to be determined self-consistently by an iterative procedure in order to ensure particle balance with the ionization rate, which in return depends on the distribution function. Indeed, in very much the same way the plasma potential, respectively the floating potential at the wall, establishes itself also in a real plasma. Details are given in section 4.

As outline above, in the Fokker-Planck operator only the diffusion term in velocity space is of relevance in connection with the isotropic distribution function $f_0(v)$ and there only the diagonal elements matter:

$$\frac{\partial f_0}{\partial t}\bigg|_{FFPO} = \left\langle \frac{1}{2}\frac{\partial^2}{\partial v_i^2}\left(\nu\left\langle \Delta v_i^2\right\rangle_\chi f_0\right)\right\rangle_{V,\Omega} = \left\langle \frac{\partial^2}{\partial v_i^2}\left(h\,f_0\right)\right\rangle_\Omega. \qquad (8)$$

The square of the velocity change $\Delta v_i$ in the direction $i$ is averaged over the corresponding stochastic parameters $\chi = \chi_1, \chi_2, \dots$ . The collision frequency $\nu$ is the relevant interaction frequency. The first step is in calculating the power function $h$:

$$h = \frac{\nu}{2}\left\langle \Delta v_i^2 \right\rangle_{\chi,V} \qquad (9)$$

The name "power function" is chosen since $m\,h$ has indeed the unit of power. The connection to the power delivered to the plasma will actually become even more obvious in section 2.3. In a second step the average over the solid angle $\Omega$ is performed, which involves also the directional derivative.

The electrons are subject to the force introduced by a spatially and temporally varying electric field and stochastic changes of the momentum direction by dominantly elastic collisions (collision frequency $\nu_m$) which convert the equation of motion to a Langevin equation. The free flight time between collisions $\tau$ is a stochastic parameter with a normalized exponential distribution $P_{\nu_m}(\tau)$:

$$P_{\nu_m}(\tau) = \nu_m\,e^{-\nu_m\tau}. \qquad (10)$$

Naturally, the interaction frequency in the Fokker-Planck operator is $\nu = \nu_m$. It should be noted already here that the above convenient distribution requires that over the free flight time $\tau$ the collision frequency $\nu_m$ is constant. Since generally the collision frequency is a function of velocity and the velocity is changing within the interval $\tau$, this requirement can be met only approximately. In section 3.4 the effect is discussed in detail and a validity condition derived. The conclusion is that the related error is probably small for most if not all relevant cases.

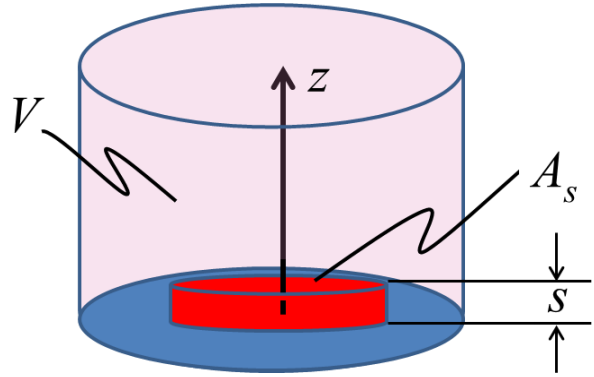

FIG. 2. Scheme of the spatial distribution of the plasma and the heating zone where stochastic heating takes place: $V$ plasma volume, $A_s$ cross section of the heating zone, $s$ thickness of the heating zone. Note that the width $s$ is not representing a hard border but is understood as a characteristic length scale. The vertical arrow indicates the $z$ axis with origin at the chamber wall and located in the center of the heating zone. While the heating zone is located in most cases at the wall as depicted in the figure, it can also appear as a localized region within the plasma as is the case in ECR discharges.

### C. Mean power delivered to the plasma and the Fokker-Planck operator

Bevor proceeding with the detailed calculation of the Fokker-Planck operator some integral properties should be investigated which relate to the mean power delivered to the plasma. This allows connection to already established approaches in the literature that do not use the Fokker-Planck operator but come to the very same result. In order to calculate the volume averaged power, or more precisely the power density, Eq. (8) is integrated over velocity after multiplication with the kinetic energy. In the following equation and the related derivation the full velocity differential $d^3v$ is temporally split into $dv_i\, d^2v_\perp$, where $d^2v_\perp = dv_j\, dv_k$ denotes the two velocity components perpendicular to $v_i$ and $i, j, k = x, y, z$.

$$
\begin{aligned}
\frac{\partial P}{\partial V} &= 4\pi \int \frac{m}{2} v^2 \frac{\partial f}{\partial t}\bigg|_{FFPO} v^2\, dv \\
&= 4\pi \int \frac{m}{2} v^2 \left\langle \frac{1}{2} \frac{\partial^2}{\partial v_i^2}\left(\nu \left\langle \Delta v_i^2 \right\rangle_\chi f_0\right)\right\rangle_{V,\Omega} v^2\, dv \\
&= \left\langle \int \frac{m}{2} v^2 \frac{1}{2} \frac{\partial^2}{\partial v_i^2}\left(\nu \left\langle \Delta v_i^2 \right\rangle_\chi f_0\right) dv^3 \right\rangle_V \\
&= \left\langle \int \frac{m}{2} v^2 \frac{1}{2} \frac{\partial^2}{\partial v_i^2}\left(\nu \left\langle \Delta v_i^2 \right\rangle_\chi f_0\right) dv_i\, dv_\perp^2 \right\rangle_V \\
&= \left\langle \int \frac{m}{2} \nu \left\langle \Delta v_i^2 \right\rangle_\chi f_0\, dv^3 \right\rangle_V \\
&= \int m\, h\, f_0\, dv^3 .
\end{aligned}
\tag{11}
$$

When applying the result for the special case of collisionless heating in inductively coupled plasmas (ICP) one has to define an effective collision frequency of the plasma electrons with the heating zone. The basic idea is therefore to describe the interaction with the field as an effective collision process (Fig. 2). The heating zone is viewed as a macro particle with a cross section $A_s$, an extension $s$ into the plasma ($z$ coordinate is perpendicular to the surface), and a density $n_s = 1/V$ with the plasma of volume $V$. The collision rate of electrons of velocity $v_z$ with the heating zone is then given by:

$$\nu\left(\vec{v}\right) = \vec{A}_s \cdot \vec{v}/V\,\Theta\left(-v_z\right) = \left(\left|v_z\right|A_s/V\right)\Theta\left(-v_z\right) = \left(\left|v_z\right|/L\right)\Theta\left(-v_z\right). \tag{12}$$

The Heaviside theta function $\Theta\left(-v_z\right)$ ensures that only electrons flowing towards the surface are considered. The characteristic length scale of the plasma is $L = V/A_s$. If $L \gg s$ the interaction time with the heating zone $\tau_s = 2s/\left|v_z\right|$ is much shorter than the time between collisions, i.e. interactions, $\tau = L/\left|v_z\right|$. In this way the interaction with the field is indeed not much different from a collisional interaction with particles. However, instead of an average loss of energy here energy is gained by the electron. Eq. (11) now reads:

$$\frac{\partial P}{\partial V} = = \int \frac{\left|v_z\right|}{L}\frac{m}{2}\left\langle \Delta v_i^2\right\rangle_\chi f_0\,\theta\left(-v_z\right)dv^3. \tag{13}$$

As is shown below this is the identical expression used by Lieberman for the calculation of the power per area delivered to the electrons in ICP (and ECR) discharges and by Czarnetzki for the INCA discharge [23], [30], [29]. There the equation is derived from the balance of the energy flux across the imaginary surface $A_s$ separating the plasma and the heating zone. It is assumed that due to the finite penetration length of the field $s$ (skin depth), heating occurs only in the region between this surface and the dielectric surface of the antenna where electrons are reflected. The energy flux in $z$-direction across the surface is split into two parts: a) electrons leaving the plasma and b) electrons returning to the plasma after interacting with the field in the heating region. The former part is:

$$\left.\frac{\partial P}{\partial A}\right|_- = \int v_z \frac{m}{2}v^2 f_0\left(v\right)\theta\left(-v_z\right)dv^3. \tag{14}$$

Note that the sign of this expression is negative since the plasma is losing energy. This flux is reflected from the surface at the dielectric window in front of the antenna and returns back to the plasma. However, while the electrons transverse the heating zone they interact with the electric field and their kinetic energy is altered by change of a velocity component $\Delta\vec{v}_i$ perpendicular to the flow direction . The sign is now positive since energy is flowing into the plasma.

$$\left.\frac{\partial P}{\partial A}\right|_+ = \int v_z \frac{m}{2}\left(\vec{v}+\Delta\vec{v}_i\right)^2 f_0\left(v\right)\theta\left(v_z\right)dv^3. \tag{15}$$

This expression needs to be averaged over the stochastic parameters of the interaction $\chi$ which includes in particular an average over the phase of the harmonic electric field. This latter average immediately causes the mean value of the velocity change to vanish:

$$\left\langle \left. \frac{\partial P}{\partial A} \right|_{+} \right\rangle = \int v_z \frac{m}{2} \left\langle (\vec{v} + \Delta \vec{v}_i)^2 \right\rangle_{\chi} f_0(v)\, \theta(v_z)\, dv^3$$
$$= \int v_z \frac{m}{2} \left( v^2 + \left\langle \Delta \vec{v}_i^2 \right\rangle_{\chi} \right) f_0(v)\, \theta(v_z)\, dv^3. \tag{16}$$

The sum of these two fluxes provides the power per area delivered from the heating zone to the plasma. The total power follows by integrating over the cross section of the area, which is identical to a multiplication by the area $A_s$ for a homogeneous field. For an inhomogeneous field, the field amplitude $\vec{E}_0$ might be understood as the root-mean square value since $\Delta \vec{v} \sim \vec{E}_0$ in any case. The average power per volume delivered to the total plasma results by division by the plasma volume $V$. Apparently, the combination of these two operations is identical to dividing by the characteristic length $L = V / A_s$:

$$\frac{\partial P}{\partial V} = \frac{1}{L} \left( \frac{\partial P}{\partial A}_{+} + \left\langle \frac{\partial P}{\partial A} \right\rangle_{-} \right) = \int \frac{|v_z|}{L} \frac{m}{2} \left\langle \Delta v_i^2 \right\rangle_{\chi} f_0(v) \theta(-v_z) dv^3. \tag{17}$$

In conclusion, the integral expressions for the power delivered to the plasma resulting from the two alternative approaches are identical. This is no guarantee that also the detailed values from the full Boltzmann equation and the Fokker-Planck equation are identical but it is certainly a strong argument in favor of using the alternative approach. In the following section it will be shown that the effective collision frequency with the field region for the collisionless case results also naturally from a more general description without invoking a physical picture as done above.

## III. DERIVATION OF THE FOKKER-PLANCK HEATING OPERATOR

### A. Fourier transformation calculation of the power function $h$

In order to calculate the power function $h$ defined by Eq. (9) it is necessary to calculate the velocity change $\Delta \vec{v}$ which again requires integration of the equation of motion for a time and space dependent electric field. The parallel action of a magnetic field and the related Lorentz force will be the subject of future work, especially in order to extend application of the Fokker-Planck concept to ECR discharges. The essential idea is now to express a general electric field of arbitrary spatial and temporal structure by a Fourier integral. Formally, it should be a Fourier series since the size of the plasma is in fact finite. However, since the length scale can be assumed to be much larger than any gradient length related to the heating field, transition to the continuum can be safely made without introducing a significant error. Then the general form of the field reads:

$$\vec{E}\left(\vec{r},t\right)=\frac{1}{\left(2\pi\right)^{2}}\iint\hat{\vec{E}}\left(\vec{k},\omega\right)e^{i\left(\vec{k}\cdot\vec{r}+\omega t\right)}d^{3}k\,d\omega. \tag{18}$$

The first integral sign represents the three-dimensional wavenumber integration and the second represents frequency integration. The advantage of this approach is that all averages can be carried out generally so that in the end only the integral over $k$ remains. Further, the analysis can be carried out without specifying the spatial-temporal structure of the field. Probably most importantly, there is no need to make any specific assumptions about collisional local or collisionless non-local heating. Both effects are inherently included by this approach.

The first step is in integrating the equation of motion in order to calculate the velocity change within a collisonless time interval $\tau$ :

$$\Delta\vec{v}=-\frac{e}{m}\int\limits_{0}^{\tau}\vec{E}\left(\vec{r},t+t_{0}\right)dt. \tag{19}$$

The spatial coordinate on the right hand side is expressed by the unperturbed motion: $\vec{r}=\vec{r}_{0}+\vec{v}\,t$. This is actually exact if $\vec{E}\perp\vec{k}$ which applies in particular for ICPs and ECRs. In cases where $\vec{E}\parallel\vec{k}$, this is equivalent to a first order approximation, as was demonstrated for the complex field structure in the INCA discharge [29]. Introduction of the time $t_{0}$ allows for an arbitrary phase $\omega t_{0}$ of the electric field for a particle starting at a position $\vec{r}_{0}$ at time $t=0$. Then the velocity change reads:

$$\begin{aligned}\Delta\vec{v}&=-\frac{e}{m}\frac{1}{\left(2\pi\right)^{2}}\iint\hat{\vec{E}}\left(\vec{k},\omega\right)e^{i\left(\vec{k}\cdot\vec{r}_{0}+\omega t_{0}\right)}\int\limits_{0}^{\tau}e^{i\left(\vec{k}\cdot\vec{v}+\omega\right)t}\,dt\,d^{3}k\,d\omega\\&=-\frac{e}{m}\frac{1}{\left(2\pi\right)^{2}}\iint\hat{\vec{E}}\left(\vec{k},\omega\right)e^{i\left(\vec{k}\cdot\vec{r}_{0}+\omega t_{0}\right)}\frac{e^{i\left(\vec{k}\cdot\vec{v}+\omega\right)\tau}-1}{i\left(\vec{k}\cdot\vec{v}+\omega\right)}d^{3}k\,d\omega.\end{aligned} \tag{20}$$

The above expression contains three stochastic parameters over which it needs to be averaged: $t_{0},\vec{r}_{0},\tau$. Apparently the average over $t_{0}$ is zero for $\omega\neq0$, reflecting the fact that an oscillating field is not causing any average drift. However, the square of the velocity change gives a non-zero result, representing diffusion in velocity space. Averaging over the phase and the starting position can be handled in parallel since they lead to similar integrals:

$$\begin{aligned}\left\langle\Delta\vec{v}^{2}\right\rangle_{t_{0},\vec{r}_{0}}&=\left(\frac{e}{m}\right)^{2}\frac{1}{\left(2\pi\right)^{4}}\iiint\hat{\vec{E}}\left(\vec{k},\omega\right)\cdot\hat{\vec{E}}\left(\vec{k}\,',\omega'\right)\frac{e^{i\left(\vec{k}\cdot\vec{v}+\omega\right)\tau}-1}{i\left(\vec{k}\cdot\vec{v}+\omega\right)}\frac{e^{i\left(\vec{k}\,'\cdot\vec{v}+\omega'\right)\tau}-1}{i\left(\vec{k}\,'\cdot\vec{v}+\omega'\right)}\\&\quad\times\psi\left(\vec{k}+\vec{k}\,',\omega+\omega'\right)d^{3}k\,d\omega\,d^{3}k\,'d\omega',\end{aligned} \tag{21}$$

with

$$\psi\left(\vec{k}+\vec{k}\,',\omega+\omega'\right) = \lim_{V,T\to\infty}\frac{1}{V\,T}\int\limits_{-T/2}^{T/2}\int e^{i\left(\left(\vec{k}+\vec{k}\,'\right)\cdot\vec{r}_0+(\omega+\omega')t_0\right)}d^3r_0\,dt_0$$
$$= \lim_{V,T\to\infty}\prod_{i=x,y,z}\mathrm{sinc}\left(\frac{k_i+k_i{}'}{2}L_i\right)\mathrm{sinc}\left(\frac{\omega+\omega'}{2}T\right). \tag{22}$$

This series of nested integrals collapses immediately to a much simpler expression since $\psi$ effectively converges to a product of delta-functions. However, the analysis of the action of $\psi$ has to be carried out with care. Firstly, the length scales of the plasma $L_i$ are large, i.e. $\left|L_i k_i\right|\gg 1$, but finite. Therefore, use is made of the approximate convergence to a delta-function but still the volume $V=L_x L_y L_z$ is treated as finite and kept explicitly in the remaining expression. Then $\psi$ reads:

$$\psi\left(\vec{k}+\vec{k}\,',\omega+\omega'\right) = \lim_{T\to\infty}\mathrm{sinc}\left(\frac{\omega+\omega'}{2}T\right)\frac{(2\pi)^3}{V}\delta\left(\vec{k}+\vec{k}\,'\right). \tag{23}$$

The same argument applies in general also to the time interval $T$ over which $t_0$ varies, although there is no real physical limitation to $T$ and the argument of extending its value to infinity is even more justified than in case of the system size. Nevertheless, here the situation is a bit more subtle since in most cases the electric fields oscillate only at a single discrete frequency $\omega_0$ so that the Fourier amplitudes are delta-functions in the frequency domain at $\omega=\pm\omega_0$:

$$\hat{\vec{E}}\left(\vec{k},\omega\right) = \hat{\vec{E}}_0\left(\vec{k}\right)\sqrt{\frac{\pi}{2}}\left(\delta\left(\omega+\omega_0\right)+\delta\left(\omega-\omega_0\right)\right) \tag{24}$$

Since Eq. (21) contains products of the amplitudes, products of delta-functions result and the further treatment requires some caution. Naturally, problems are avoided if the delta-functions in the frequency domain are evaluated first. Then, the sinc-function in the frequency domain enforces in the $\omega,\omega'$ integration that only frequency pairs of opposite sign give a non-zero result for $\psi$, i.e. yielding a factor of 1. The argument can be extended to electric field composed of a series of discrete frequencies $\omega_0,\omega_1,...$ with amplitudes $\hat{\vec{E}}_0\left(\vec{k}\right),\hat{\vec{E}}_1\left(\vec{k}\right),....$ Clearly, pairs of different frequencies give zero contribution. It is therefore sufficient and also more convenient to continue the analysis from here for a single frequency $\omega_0$ but keeping in mind that the final result can always be extended to more frequencies by a simple summation. Integration with respect to $k'$ becomes straight forward since now the delta-function resulting from the sinc-function (Eq. (23)) can be evaluated:

$$\left\langle\Delta\vec{v}^2\right\rangle_{t_0,\vec{r}_0} = \frac{1}{2V}\left(\frac{e}{m}\right)^2\int\left|\hat{\vec{E}}_0\left(\vec{k}\right)\right|^2\left(\frac{1-\cos\left(\left(\vec{k}\cdot\vec{v}+\omega_0\right)\tau\right)}{\left(\vec{k}\cdot\vec{v}+\omega_0\right)^2}+\frac{1-\cos\left(\left(\vec{k}\cdot\vec{v}-\omega_0\right)\tau\right)}{\left(\vec{k}\cdot\vec{v}-\omega_0\right)^2}\right)d^3k. \tag{25}$$

In evaluating the delta-function integrals, advantage was taken of the fact that in order for electric field to be real, $\hat{\vec{E}}^{*}\left(\vec{k},\omega\right)=\hat{\vec{E}}\left(-\vec{k},-\omega\right)$ must apply. The final step is in averaging over the collisonless period $\tau$ which makes the connection to the momentum changing collision frequency $\nu_{m}$. Further the expression is multiplied by the rate $\nu_{m}/2$ required for the Fokker-Planck operator and the $h$-function, respectively:

$$
\begin{aligned}
\frac{\nu_{m}}{2}\left\langle \Delta \vec{v}^{2}\right\rangle_{t_{0},\vec{r}_{0},\tau} &= \frac{1}{2V}\left(\frac{e}{m}\right)^{2}\int\left|\hat{\vec{E}}_{0}\left(\vec{k}\right)\right|^{2} \\
&\times \frac{\nu_{m}}{2}\int_{0}^{\infty}\left(\frac{1-\cos\left(\left(\vec{k}\cdot\vec{v}+\omega_{0}\right)\tau\right)}{\left(\vec{k}\cdot\vec{v}+\omega_{0}\right)^{2}}+\frac{1-\cos\left(\left(\vec{k}\cdot\vec{v}-\omega_{0}\right)\tau\right)}{\left(\vec{k}\cdot\vec{v}-\omega_{0}\right)^{2}}\right)\nu_{m}\,e^{-\nu_{m}\tau}\,d\tau\,d^{3}k \quad (26)\\
&= \frac{1}{2V}\left(\frac{e}{m}\right)^{2}\int\left|\hat{\vec{E}}_{0}\left(\vec{k}\right)\right|^{2}\,K\left(\vec{k}\cdot\vec{v},\,\omega_{0},\nu_{m}\right)d^{3}k,
\end{aligned}
$$

The $h$-function finally reads:

$$
h=\frac{1}{2V}\left(\frac{e}{m}\right)^{2}\int\left|\hat{\vec{E}}_{0}\left(\vec{k}\right)\right|^{2}K\left(\vec{k}\cdot\vec{v},\,\omega_{0},\,\nu_{m}\right)d^{3}k, \qquad (27)
$$

where

$$
K\left(\vec{k}\cdot\vec{v},\,\omega_{0},\nu_{m}\right)=\frac{1}{2}\left(\frac{\nu_{m}}{\left(\vec{k}\cdot\vec{v}+\omega_{0}\right)^{2}+\nu_{m}^{2}}+\frac{\nu_{m}}{\left(\vec{k}\cdot\vec{v}-\omega_{0}\right)^{2}+\nu_{m}^{2}}\right). \qquad (28)
$$

This general result combines Ohmic as well as stochastic heating and applies for arbitrary spatially structured electric fields. The limiting cases of pure Ohmic and pure stochastic heating can be readily identified as outlined below. It might be repeated that for more than one frequency in the spectrum, the resulting $h$-function is the sum of the $h$-functions of the individual frequencies. The integral consist only of the absolute square of the Fourier amplitudes of the electric field and a kernel that has the simple form of a Lorentz function. Note that the kernel $K$ is symmetric with respect to the reversal of the sign of the scalar product $\vec{k}\cdot\vec{v}$. This kernel is named the conductivity kernel since it can be identified as the real part of the kernel of the Fourier transform of the general kinetic conductivity $\hat{\sigma}$ (for an isotropic background distribution $f_{0}\left(v\right)$ and a harmonic and inhomogeneous electric field) [31], [24]:

$$
\hat{\sigma}=\frac{e^{2}}{2m}\int\left(\frac{1}{i\left(\vec{k}\cdot\vec{v}+\omega_{0}\right)+\nu_{m}}+\frac{1}{i\left(\vec{k}\cdot\vec{v}-\omega_{0}\right)+\nu_{m}}\right)f_{0}\left(v\right)d^{3}v, \qquad (29)
$$

so that:

$$\mathrm{Re}\left(\hat{\sigma}\right) = \frac{e^2}{2m}\int\left(\frac{\nu_m}{\left(\vec{k}\cdot\vec{v}+\omega_0\right)^2+\nu_m^2}+\frac{\nu_m}{\left(\vec{k}\cdot\vec{v}-\omega_0\right)^2+\nu_m^2}\right)f_0\left(v\right)d^3v. \qquad (30)$$

The missing factor $1/m$ results from the fact that the Fokker-Planck operator contains $h = \nu\left\langle\Delta\vec{v}^2\right\rangle/2$ and not $\nu\,m\,\Delta\vec{v}^2/2$, which has the dimension of power. Recalling that the Fourier transform of the power density is in general $\mathrm{Re}\left(\hat{\sigma}\right)\left|\hat{\tilde{E}}\right|^2/2$, the final result found here is rather obvious in retrospect.

A convenient consequence of the above form of $h$ is that both outstanding operations for the global and isotropic Fokker-Planck operator, differentiation with respect to velocity and integration over the solid angle in velocity space, act only on the conductivity kernel and can therefore be carried out generally without specifying the particular spatial structure of the electric field.

## B. Limiting cases of pure stochastic and Ohmic heating

In general, the field can have a complicate spatial structure. However, for all practical cases the spatial variation responsible for stochastic heating is mainly in one particular direction and a much weaker variation is found in the perpendicular direction. For instance, in an ICP the direction of the evanescent penetration of the field into the plasma has usually the shortest length scale. Additionally the field strength usually varies also perpendicularly across the plain of the antenna but on a much larger length scale. Similar conditions with a dominant direction of spatial variation are found also for ECR or INCA discharges. If the major direction of spatial variation of the electric field is denoted as $\parallel$ and the perpendicular plane as $\perp$, then $\left|\vec{k}_\perp\right| << \left|\vec{k}_\parallel\right|$. Therefore, in the conductivity kernel the approximation can be made $\vec{k}\cdot\vec{v} = k_\parallel\,v_\parallel + \vec{k}_\perp\cdot\vec{v} \approx k_\parallel\,v_\parallel$ since the velocity distribution is isotropic. This allows introduction of new dimensionless coordinates in the vicinity of the respective resonances of the conductivity kernel: $\kappa_\pm = \left(k_\parallel\,v_\parallel \pm \omega_0\right)/\nu_m$. Integration can now be split into parallel and perpendicular directions. Since the integral form for the two resonances is identical, a common variable $\kappa$ can be introduced:

$$h = \frac{1}{4V}\left(\frac{e}{m}\right)^2\frac{1}{v_\parallel}\iint\left(\left|\hat{\tilde{E}}_0\left(\frac{\kappa\nu_m-\omega_0}{v_\parallel},\vec{k}_\perp\right)\right|^2+\left|\hat{\tilde{E}}_0\left(\frac{\kappa\nu_m+\omega_0}{v_\parallel},\vec{k}_\perp\right)\right|^2\right)\frac{1}{1+\kappa^2}\,d\kappa\,d^2k_\perp. \qquad (31)$$

In the new normalized coordinate $\kappa$ the conductivity kernel has significant values only within a range of one. Then it is obvious that the Fourier amplitudes become effectively independent of $\kappa$ if $\nu_m << \omega_0$. However, this is meaningful only if the Fourier amplitudes have significant values at $\hat{\vec{E}}\left(\pm\dfrac{\omega_0}{v_{\|}},\vec{k}_\perp\right)$. This is the case if the relevant length scale of the spatial variation $\ell$, which defines the $k$ spectrum, is sufficiently short, i.e. $\ell\,\omega_0 < v_{\|} \approx v_{th}$, where the characteristic parallel velocity is identified as the thermal velocity of the electrons. The physical reason is that the electrons must pass thermally through the inhomogeneous field region within a time shorter than the RF period in order to avoid the inefficient local quiver motion. Under these conditions and using the symmetry of the Fourier transform of the electric field for positive and negative frequencies the following approximation can be made:

$$
\begin{aligned}
h &\approx \frac{1}{4V}\left(\frac{e}{m}\right)^2\frac{1}{v_{\|}}\int\left(\left|\hat{\vec{E}}_0\left(\frac{\omega_0}{v_{\|}},\vec{k}_\perp\right)\right|^2+\left|\hat{\vec{E}}_0\left(-\frac{\omega_0}{v_{\|}},\vec{k}_\perp\right)\right|^2\right)d^2k_\perp\int_{-\infty}^{\infty}\frac{1}{1+\kappa^2}\,d\kappa \\
&= \frac{\pi}{2V}\left(\frac{e}{m}\right)^2\frac{1}{v_{\|}}\int\left|\hat{\vec{E}}_0\left(\frac{\omega_0}{v_{\|}},\vec{k}_\perp\right)\right|^2 d^2k_\perp \\
&= \frac{\pi A_\perp}{2V}\left(\frac{e}{m}\right)^2\frac{1}{v_{\|}}\left\langle\left|\hat{\vec{E}}_0\left(\frac{\omega_0}{v_{\|}}\right)\right|^2\right\rangle_{A_\perp}.
\end{aligned}
\tag{32}
$$

By Parseval's theorem integration over the perpendicular wavenumbers is equivalent to integration over the corresponding area $A_\perp$ which leads to the last line with the area averaged squared electric field [43]: $\int\left|\hat{\vec{E}}_0\left(\omega_0/v_{\|},\vec{k}_\perp\right)\right|^2 d^2k_\perp = \int\left|\vec{E}_0\left(\omega_0/v_{\|},\vec{r}_\perp\right)\right|^2 d^2r_\perp$. The notation in the above formula is meant to represent the absolute square of the amplitude of the one-dimensional Fourier transform (in the direction of the dominant spatial variation) of the electric field at the specific wavenumber of $k_{\|} = \omega_0/v_{\|}$.

One of the main insights from the above result (Eq. (32)) is that stochastic heating in general scales like $1/v_{\|}$ times a factor that depends on the spatial profile and the frequency spectrum of the electric field. It should be noted that the term $1/v_{\|}$ is not causing a divergence for $v_{\|}\to 0$ under any circumstances. In order to be a proper behaving Fourier amplitude, $\hat{\vec{E}}$ must at least scale like $k_{\|}^{-(1+\alpha)}$ with $\alpha > 0$. Therefore, $\left|\hat{\vec{E}}\left(\omega_0/v_{\|}\right)\right|^2$ scales at least like $v_{\|}^{2(1+\alpha)}$ which leads generally to $\lim_{v_{\|}\to 0} h = 0$. Further, the same result applies also at least to the first derivative with respect to $v_{\|}$.

The opposite limit of Ohmic heating applies if at least one of the two following conditions holds: $v_m \gg \omega_0$ or $\ell\,\omega_0 \gg v_{\|}\approx v_{th}$. Then the term $\vec{k}\cdot\vec{v}$ can be neglected in the interaction kernel which allows the term to be moved out of the integral:

$$h = \frac{1}{2V} \left(\frac{e}{m}\right)^2 \frac{\nu_m}{\omega_0^2 + \nu_m^2} \int \left|\hat{\vec{E}}_0\left(\vec{k}\right)\right|^2 d^3k \tag{33}$$

Identifying the volume average of the squared field in the above expression as the area average by using again Parseval's theorem [43], this expression reduces to the well-known local result:

$$h = \left(\frac{e}{m}\right)^2 \frac{1}{2} \frac{\nu_m}{\omega_0^2 + \nu_m^2} \left\langle \left|\vec{E}_0\right|^2 \right\rangle_V = \left\langle \frac{\vec{j} \cdot \vec{E}}{m\, n_e} \right\rangle_{V,t} \tag{34}$$

The second expression follows by identifying the classical Ohmic conductivity and denoting by $n_e$ the electron density. The factor ½ in the first expression can be interpreted to be the result of the temporal average of the oscillating field squared. Therefore, this special case is fully confirming the above more general interpretation in terms of the conductivity.

### C. Example: Classical ICP with exponentially decaying field

It is now straight forward to analyze a particular field distribution as an example. In case of a standard ICP plasma the electric field varies harmonically at a frequency $\omega_0$ and the field decays exponentially with distance from the antenna (dielectric window) $\sim \exp\left(-z/s\right)$ with $s$ denoting the skin depth [30]. The electric field enters the plasma at the origin of the coordinate system and it is polarized perpendicular to the direction of the spatial decay, e.g. in $x$-direction. In order to account for particle reflection at the surface, the field is mirrored at the origin. Since this artificially doubles the plasma volume, a factor ½ needs to be introduced in the $h$-function which scales like $1/V$. With $k_\parallel = k_z$ the spatial Fourier transform reads:

$$\hat{\vec{E}} = \vec{E}_0 \frac{1}{\sqrt{2\pi}} \int_{-\infty}^{\infty} e^{-\left|\frac{z}{s}\right|} e^{ik_z z}\, dz = \vec{E}_0 \sqrt{\frac{2}{\pi}} \frac{s}{1 + \left(k\, s\right)^2}. \tag{35}$$

Inserting this expression in Eq. (32) gives the stochastic expression for the case of negligible collisonality and sufficiently large field gradient. Please note that here $\vec{k} \cdot \vec{v} = \left|\vec{k}\right|\left|\vec{v}\right|\cos\left(\vartheta\right) = \left|k_z\right| v_z$, where $\cos\left(\vartheta\right)$ is the angle to the $z$-axis.

$$h = \frac{A_\perp\, s^2}{2V} \left(\frac{e}{m}\right)^2 \left\langle \left|\vec{E}_0\right|^2 \right\rangle_{A_\perp} \frac{\left|v_z^3\right|}{\left(v_z^2 + \left(\omega_0\, s\right)^2\right)^2}. \tag{36}$$

This is exactly the expression derived first by Lieberman in connection with calculating the integral for the mean power delivered to the plasma (Eq. (11) or (13), respectively) if $v_\parallel$ is identified as $v_z$ [23], [30]. Further, the effective collision frequency $\nu = \left|v_z\right|/L = \left|v_z\right| A_\perp / V$ introduced by physical arguments in Eq. (12) results naturally here in the collisionless limit. A

more general expression combining stochastic and Ohmic heating is derived by returning to Eq. (27).

$$
\begin{aligned}
h &= \frac{A_\perp s^2}{2\pi V}\left(\frac{e}{m}\right)^2 \left\langle \left|\vec{E}_0\right|^2\right\rangle_{A_\perp} \int_{-\infty}^{\infty} \frac{1}{\left(1+\left(k_z s\right)^2\right)^2} \mathrm{K}\left(\left|k_z\right|v_z,\omega_0,\nu_m\right)dk_z \\
&= \frac{A_\perp s^2}{2V}\left(\frac{e}{m}\right)^2 \left\langle \left|\vec{E}_0\right|^2\right\rangle_{A_\perp} \left(\frac{\left|v_z\right|\left(\left|v_z\right|+s\nu_m\right)^2}{\left(\left(s\omega_0\right)^2+\left(\left|v_z\right|+s\nu_m\right)^2\right)^2}+\frac{1}{2}\frac{s\nu_m}{\left(s\omega_0\right)^2+\left(\left|v_z\right|+s\nu_m\right)^2}\right).
\end{aligned}
\tag{37}
$$

The first term describes stochastic heating in case of vanishing collisionality or a very strong spatial gradient $\left(s\nu_m \to 0\right)$ and the second term describes Ohmic heating in the highly collisional case or for a vanishing spatial gradient $\left(s\nu_m \gg v_{th}\right)$. Naturally, the expression converges to the limiting cases discussed above, where in case of infinite skin depth $\left(s \to \infty\right)$ the pure Ohmic case is recovered. Therefore, the *h*-function and thereby also the Fokker-Planck operator provide a smooth transition between the two relevant stochastic collisional interactions: On one hand locally with neutrals which leads to Ohmic heating and on the other hand globally with the heating zone which leads to stochastic heating. The former is included by averaging over the free flight time distribution function (Eq. (10)) and the latter by the volume averaging.

In this context it should be noted that the volume ratio $A_\perp s/\left(2V\right)$ results from volume averaging the square of the exponentially decaying field, i.e. $\int_0^\infty \left(e^{-z/s}\right)^2 dz = \frac{s}{2}$. The factor ½ in front of the Ohmic term can be interpreted as the result of the temporal average of the harmonic field squared.

The general behavior of the *h*-function is shown in Fig. 3. For the presentation normalized values and variables are used. The volume ratio discussed above is explicitly included in the amplitude factor $h_0$ by introducing a volume averaged free oscillating amplitude $\left\langle v_E^2\right\rangle_V$:

$$
v_s = s\omega, \ \ h_0 = \omega_0 \frac{A_\perp s}{2V}\left\langle \left|\frac{e\vec{E}_0}{m\omega_0}\right|^2\right\rangle_{A_\perp} = \omega_0\left\langle v_E^2\right\rangle_V.
\tag{38}
$$

The shape looks very symmetric with respect to the dependence on the two normalized variables $\nu_m/\omega_0$ and $v_z/v_s = v_z/\left(s\omega_0\right)$. For static electrons $\left(v_z = 0\right)$ or alternatively a homogeneous field $\left(s \to \infty\right)$ the typical shape of the Ohmic function is exhibited. For vanishing collisionality $\left(\nu_m \to 0\right)$ very much the same shape is found along the velocity axes, supporting the interpretation of $v/s$ as an effective stochastic collision frequency. However it should be noted that the peak value of the curve along the velocity axes is slightly higher than the corresponding value along the collision axis by a factor $3^{3/2}/4 = 1.30$. The effect of low collisionality $\nu_m/\omega_0 < 1$ is small throughout except for zero or close to zero velocity, since there stochastic heating vanishes. At high collisonality the function converges generally to a homogeneous low value defined by the collision frequency alone.

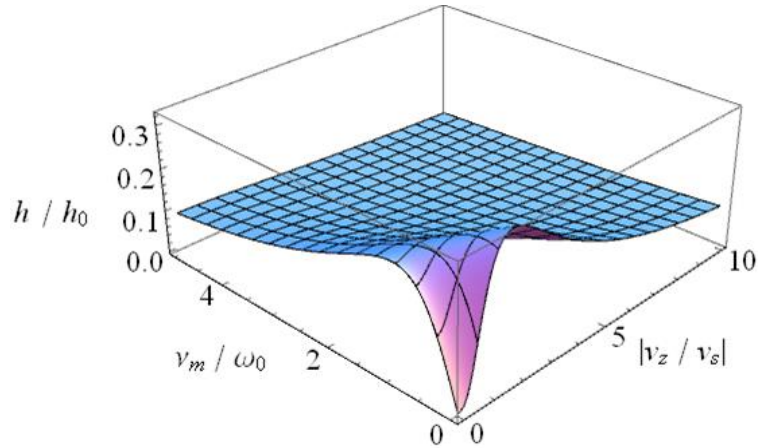

FIG. 3. The power *h*-function for an ICP (Eq. (37)) as a function of velocity and collision frequency as independent variables. Normalization parameters are defined in Eq. (38).

Vahedi et al. have also attempted to include collisions in their integral calculation of the power deposition in ICPs [30]. However, their approach to calculate $\langle \Delta \vec{v}^2 \rangle$ is inconsistent. Elastic collisions are treated as a friction term like in fluid dynamics. In order to avoid that friction in *z*-direction actually stops electrons from flowing in and out of the heating zone (skin layer), they allow friction to act only in the perpendicular direction of the electric field. Naturally, their general result is not fully correct, although the model converges to the correct limits in cases of zero and infinite collisionality.

### D. Solid angle average and differentiation of the *h* function

The average over the solid angle involves the *h*-function as well as the directional derivatives which act on *h* and the isotropic distribution function $f_0(v)$. It is obvious that after averaging all derivatives will appear only with respect to the absolute velocity *v* and the operator will remain to be of second order. Further, in *h* all operations act only on the conductivity kernel defined by Eq. (28). Since the velocity component in the derivatives points in the direction of the electric field $\vec{E}$ but the velocity component in the conductivity kernel points in the direction of $\vec{k}$, i.e. the direction in which $\vec{E}$ varies, two general cases can be distinguished:

a) $\vec{k} \perp \vec{E}$ which is the case for ICPs, ECRs, and also the new INCA discharge. This is probably the more important and certainly the most straight forward case. However, in order to treat ECR discharges properly, the $h$ function still needs to be calculated using the Lorentz force and with a static magnetic field perpendicular to both, $\vec{k}$ and $\vec{E}$.

b) $\vec{k} \parallel \vec{E}$, which is the case for instance in CCPs. However, the spatial and temporal field structure in CCPs is complicate and the correct treatment of the electron kinetics and the field is a challenge in itself. Clearly this goes beyond the scope of this work. Note that in this case space charge can play an important role and Poisson's equation needs to be solved in general. This case includes also Landau damping. Like for any wave related field structure, $k$ and $\omega$ are no longer independent variables but are coupled by a certain dispersion relation. This second case b) will not be further investigated in this work, although one can follow generally the same concept as derived here for case a) here.

The solid angle averaging will be applied directly to the conductivity kernel $K$. However, for practical purposes of analytical integration it might be easier in some cases to carry out the Fourier integral over $k$ first. This motivates to analyze the general action of the derivatives and solid angle averaging on $h$. The result can then be applied either prior to the $k$-integration with action on $K$ or subsequently on the integrated $h$-function. Further, depending on the choice of the spatial electric field profile an analytical solution of the final $d^3k$ might not exist. Nevertheless, for the intended use in a Boltzmann solver, which integrates numerically anyway, a numerical table of the final coupling function (Eq. (42) below) would be as good as an analytical expression or even better. In Appendix B the details of the solid angle averaging and the combination with the derivatives are derived. The result can be identified as a generalization of the local two-term heating operator (Eq. (3)):

$$\left.\frac{\partial f}{\partial t}\right|_{FFPO} = \frac{\omega_0 \left\langle v_E^2 \right\rangle_V}{6v^2} \frac{\partial}{\partial v}\left( g_1(v) v f_0 + g_2(v) v^2 \frac{\partial}{\partial v} f_0 \right). \tag{39}$$

The dimensionless coupling functions $g_{1,2}(v)$ are independent of the electric field amplitude due to the normalization by $h_0$. Note that this normalization leads to a global generalization of the local factor $v_E^2$ in the two-term heating operator (Eq. (3)). The same normalization was used already in the presentation of the $h$-function for the special case of an ICP. Further, it should be recalled that in cases where the electric field contains more than a single frequency, the corresponding Fokker-Planck operators for each frequency add linearly.

$$g_1(v, \nu_m) = v \frac{\partial \nu_m}{\partial v} \frac{\partial g_2}{\partial \nu_m} = \left( 1 + \frac{v}{\sigma_m} \frac{\partial \sigma_m}{\partial v} \right) \nu_m \frac{\partial g_2}{\partial \nu_m},$$

$$g_2(v, \nu_m) = \frac{3}{2} \int_{-1}^{1} \left( 1 - \xi^2 \right) \frac{h(v\xi, \nu_m)}{h_0} d\xi, \tag{40}$$

$$h_0 = \omega_0 \left\langle v_E^2 \right\rangle_V = \omega_0 \left\langle \left| \frac{e\vec{E}_0}{m\,\omega_0} \right|^2 \right\rangle_V.$$

Here again $v_z = v \cos(\vartheta) = v \xi$ was used. Formally it is sufficient to calculate the function $g_2$, which then allows to derive $g_1$ by differentiating with respect to $\nu_m = N_g v \sigma_m(v)$, where $\sigma_m$ is the effective cross section for momentum changing collisions and $N_g$ the gas density. If the operation is applied directly to the conductivity kernel, one can define related dimensionless functions $\gamma_1$ and $\gamma_2$ (using $k_\parallel = k_z$):

$$\gamma_1\left(k_\parallel \, v, \omega_0, \nu_m\right) = \nu_m \frac{\partial \gamma_2}{\partial \nu_m} = \omega_0 \frac{3}{2} \int_{-1}^{1} \left(1 - \xi^2\right) \nu_m \frac{\partial}{\partial \nu_m} \mathrm{K}\left(|k_z| \, v \, \xi, \omega_0, \nu_m\right) d\xi,$$

$$\gamma_2\left(k_\parallel \, v, \omega_0, \nu_m\right) = \omega_0 \frac{3}{2} \int_{-1}^{1} \left(1 - \xi^2\right) \mathrm{K}\left(|k_z| \, v \, \xi, \omega_0, \nu_m\right) d\xi,$$

(41)

so that

$$g_{1,2}\left(v, \nu_m\right) = \frac{\int \left|\hat{\vec{E}}_0\left(\vec{k}\right)\right|^2 \gamma_{1,2}\left(|k_z| \, v, \omega_0, \nu_m\right) d^3k}{2 \int \left|\hat{\vec{E}}_0\left(\vec{k}\right)\right|^2 d^3k}.$$

(42)

Here the identity $\int \left|\vec{E}_0\right|^2 d^3r = \int \left|\hat{\vec{E}}_0\right|^2 d^3k$ was applied. For the case $\vec{k} \perp \vec{E}$ the general result is:

$$\gamma_1 = \frac{3\nu_m \omega_0}{2k_z^3 v^3}\left(-4k_z \, v + 2\nu_m\left(\arctan\left(\frac{k_z \, v + \omega_0}{\nu_m}\right) + \arctan\left(\frac{k_z \, v - \omega_0}{\nu_m}\right)\right) + \ln\left(\frac{\left(k_z \, v + \omega_0\right)^2 + \nu_m^2}{\left(k_z \, v - \omega_0\right)^2 + \nu_m^2}\right)\right),$$

$$\gamma_2 = \frac{3\omega_0}{2k_z^3 v^3}\left(-k_z \, v \, \nu_m + \left(k_z^2 \, v^2 + \nu_m^2 - \omega_0^2\right)\left(\arctan\left(\frac{k_z \, v + \omega_0}{\nu_m}\right) + \arctan\left(\frac{k_z \, v - \omega_0}{\nu_m}\right)\right)\right.$$

$$\left. + \nu_m \, \omega_0 \ln\left(\frac{\nu_m^2 + \left(k_z \, v + \omega_0\right)^2}{\nu_m^2 + \left(k_z \, v - \omega_0\right)^2}\right)\right).$$

(43)

The expressions are symmetric to the reversal of the sign of $k_z$ so that there is no need to explicitly indicate absolute values here. In the limit $k_z \to 0$ the expression converges towards the Ohmic case $\omega_0 \nu_m / \left(\nu_m^2 + \omega_0^2\right)$ and then indeed $g_2 = g_O$ (Eq. (4)). If further $\nu_m$ is constant, apparently $g_1$ vanishes. In this case the Fokker-Planck Eq. (39) is identical to the two-term approximation result Eq. (3). Differences appear only by the velocity dependence of $\nu_m$ which leads to the appearance of the additional term scaled by the coupling function $g_1$. This subtle difference is discussed in more detail in the following section 3.5. Like already mentioned in connection with calculating the *h*-function for an ICP in section 3.3, in case of electron reflection at a surface the electric field needs to be mirrored at the origin. Since this effectively doubles the plasma volume, $2V$ should be used when calculating $\left\langle v_E^2 \right\rangle_V$ from the mirrored electric field profile. Naturally, when using the original field limited to one half-sphere only, $V$ has to be used and the two alternative ways of calculating the average give identical results in any case.

Last not least application of the final result to the ICP case discussed above can be made. Carrying out the Fourier transform in Eq. (42) using (35) and (43) the coupling function reads:

$$
\begin{aligned}
g_1\left(v, \nu_m\right) &= \left(1 + \frac{v}{\sigma_m} \frac{\partial \sigma_m}{\partial v}\right) \nu_m \frac{\partial g_2}{\partial \nu_m} \\
&= \left(1 + \frac{v}{\sigma_m} \frac{\partial \sigma_m}{\partial v}\right) \frac{3 s^3 \omega_0^2 \nu_m}{v^3} \\
&\quad \times \left( \frac{v}{s\,\omega_0} \left( 4 - \frac{v\,\nu_m}{s\left(\omega_0^2 + \nu_m{}^2\right)} - \frac{\frac{v}{s}\left(\frac{v}{s} + \nu_m\right)}{\omega_0^2 + \left(\frac{v}{s} + \nu_m\right)^2} \right) \right. \\
&\quad \left. - 4 \arctan\left( \frac{\frac{v}{s}\omega_0}{\omega_0^2 + \nu_m\left(\frac{v}{s} + \nu_m\right)} \right) - \frac{2\nu_m}{\omega_0} \ln\left( \frac{\omega_0^2 + \left(\frac{v}{s} + \nu_m\right)^2}{\omega_0^2 + \nu_m^2} \right) \right),
\end{aligned}
\tag{44}
$$

$$
\begin{aligned}
g_2\left(v, \nu_m\right) &= \frac{3 s\,\omega_0}{v} \left( \left( \frac{1}{2} + \frac{s^2\left(\omega_0^2 - \nu_m^2\right)}{v^2} \right) \ln\left( \frac{\omega_0^2 + \left(\frac{v}{s} + \nu_m\right)^2}{\omega_0^2 + \nu_m^2} \right) - 1 \right) \\
&\quad + 6 \frac{s^3 \omega_0^2 \nu_m}{v^3} \left( \frac{v}{s\,\omega_0} - 2 \arctan\left( \frac{v\,\omega_0}{s\left(\omega_0^2 + \nu_m\left(\frac{v}{s} + \nu_m\right)\right)} \right) \right).
\end{aligned}
\tag{45}
$$

In the Ohmic case also this more special expression converges towards $g_O$. The overall shape of the coupling function $g_2\left(v, \nu_m\right)$ looks very much the same as the related $h\left(v_z, \nu_m\right)$ function shown in Fig. 3. More interesting is the behavior of the entire Fokker-Planck operator for a Maxwellian distribution function and a constant collision frequency, similar to the presentation in Fig. 1 for the Ohmic coupling function and the heating operator following from the two-term approximation. Naturally, under the assumption $\nu_m = \text{const.}$, the coupling function $g_1 = 0$. By assuming a Maxwellian distribution function with a certain temperature $T_e$ a new parameter is introduced: $\alpha = k_B T_e / \varepsilon_s$, where $\varepsilon_s = m v_s^2 / 2$ and $v_s = s\,\omega_0$ like before. Fig. 4 shows the energy dependence of the Fokker-Planck operator for 7 values of this electron temperature parameter: $\alpha = 0, 1, 2, 4, 8, 16, 32$.

Throughout the very same behavior of the operator is found as for the classical collisional case in the two-term approximation in Fig. 1. Low energy electrons are removed from the distribution and are reintroduced at higher energies. Depending on $\alpha$ and $\nu_m$ the zero crossing is not necessarily exactly at the mean energy $\varepsilon / k_B T_e = 3/2$ but indeed never far off. Naturally, the integral over energy is zero in any case, which confirms particle number conservation. At very low collisionality $\left( \nu_m / \omega_0 = 0.1 \right)$ a monotonous increase of the amplitude with $\alpha$ is observed (Fig. 4a). This confirms that stochastic heating depends on the thermal motion of the electrons through the heating zone.

The residual collisional contribution is shown by the curve for $\alpha = 0$, i.e. without any thermal motion, which has the lowest amplitude. At about $\alpha \approx 35$, optimum amplitudes of about $-0.16$ and $+0.04$, respectively, are reached. More important is the area of the positive (or negative) section of the curve, which depends also on the zero crossing. Under collisonless conditions a rather broad extreme at $\alpha \approx 18$ is obtained. Therefore, for a proper choice of $v_s$, i.e. the gradient length and the frequency, stochastic heating is only weakly dependent on the electron temperature and can be as efficient as collisional heating at its optimum $\left( \nu_m / \omega_0 = 1 \right)$.

Reducing the collisionality to values lower than in Fig. 4a, has little effect on the amplitude or the shape the curves if $\alpha > 10$. This confirms the standard interpretation that the thermal transit time of the electrons through the heating zone $\tau_s = 2 s / v_{th}$ should be short compared to the period $T_{RF}$ of the oscillating field so that the quiver motion is suppressed. Using $v_{th} = \sqrt{2 k_B T_e / m}$ and the value of $\alpha = 10$ this translates into $T_{RF} / \tau_s > 10$, although there is no strict threshold and transition between Ohmic and stochastic heating is clearly gradual.

With increasing collisonality (Fig. 4b, $\nu_m / \omega_0 = 0.5$), the behavior of the operator converges to the classical Ohmic case shown in Fig. 1. Shape and amplitude are now only weakly dependent on the value of $\alpha$. At even higher collisonality, $\nu_m / \omega_0 = 1$, the classical Ohmic heating case is fully recovered. The Fokker-Planck operator is then identical to the two-term operator as is exhibited by comparision between Fig. 4c and Fig. 1.

This finally confirms the full compatibility between the two operators under the assumption of a constant, i.e. velocity independent, momentum changing collision frequency. Nevertheless, in reality the collision frequency is not constant, at least not within the entire velocity range. Then the coupling function $g_1$ is non-zero. The role and the relative importance of this coupling function are discussed in the following section.

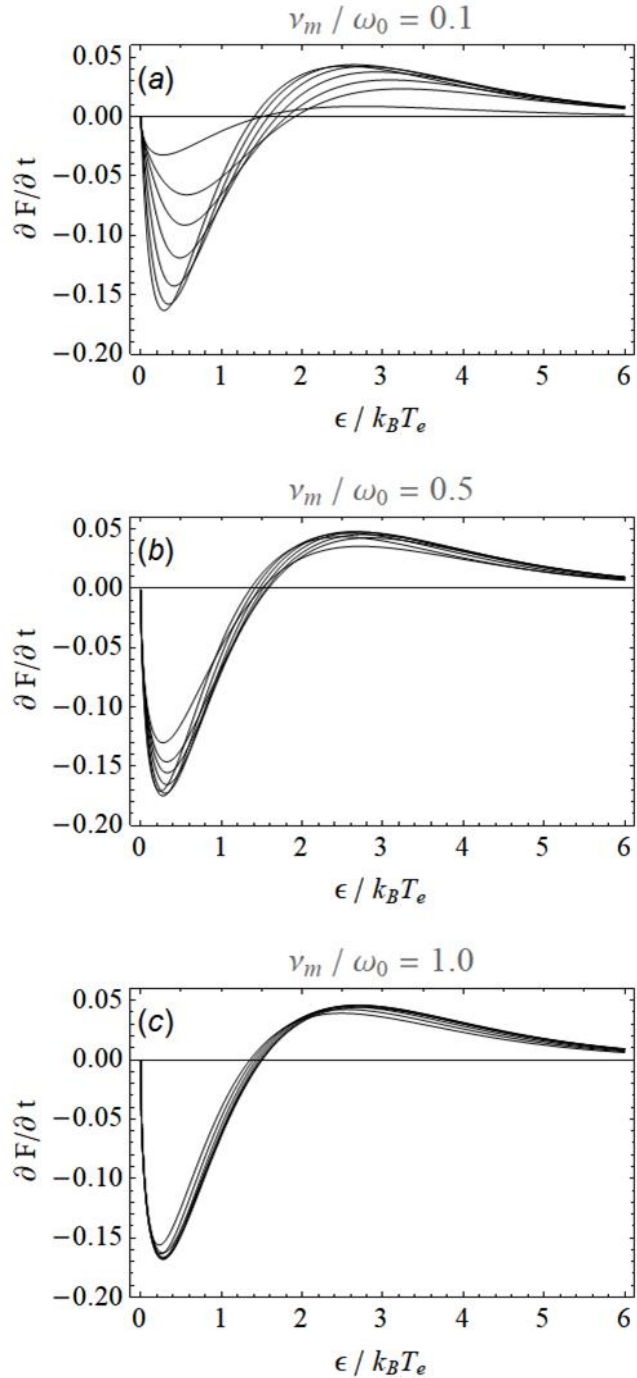

FIG. 4. Fokker-Planck heating operator for the particular case of an ICP and a Maxwell-Boltzmann distribution at an electron temperature $T_e$ and for a constant collision frequency $\nu_m$ as a function of the kinetic energy of the electrons $\varepsilon$. The relative scaling of the ordinate is the same as in Fig. 1. Each figure combines 7 curves for $\alpha = kT_e / \varepsilon_s = 0,1,2,4,8,16,32$, with the amplitude monotonously increasing with the value of $\alpha$. a) $\nu_m / \omega_0 = 0.1$, b) $\nu_m / \omega_0 = 0.5$, and c) $\nu_m / \omega_0 = 1.0$. The action of the operator is displayed for the energy distribution function $F(\varepsilon)$ and normalization is by the same factor as in Fig. 1: $m \nu_E^2 \omega_0 n_e / \sqrt{\pi}$, where $n_e$ is the electron density.

### E. Impact of the velocity dependence of the collision frequency

The above approach of including elastic collisions is strictly fully consistent only when the collision frequency is not velocity dependent. Otherwise already the probability function could not contain the simple expression $\nu_m \tau$ in the exponent, as was mentioned already when introducing the distribution function in Eq. (10). Therefore, one has to demand that $\nu_m$ should not change too strongly with velocity during the free fall time between two collisions. This demand is reflected by the presence of the coupling function $g_1$ in the final Fokker-Planck operator. This coupling function is directly proportional to the velocity derivative of the collision frequency and it is this particular term that makes a difference in the Ohmic limit to the operator resulting from two-term approximation. One can therefore demand for self-consistency that this term should make only a small if not negligible contribution in the action of the FFPO:

$$\left| g_1 f_0 \right| << \left| g_2 \, v \frac{\partial f_0}{\partial v} \right|. \tag{46}$$

Assuming without real loss in generality that the distribution function is Maxwellian, using the relation between $g_1$ and $g_2$, and switching from velocity to energy this translates into:

$$\left| \frac{k_B T_e}{g_2} \frac{\partial \nu_m}{\partial \varepsilon} \frac{\partial g_2}{\partial \nu_m} \right| << 1. \tag{47}$$

The most pronounced effect can be expected in the Ohmic limit while in the collisonless limit the question of the velocity dependence is superfluous anyway. Using therefore the Ohmic coupling function the condition becomes:

$$\left| \frac{k_B T_e}{\nu_m} \frac{\partial \nu_m}{\partial \varepsilon} \frac{\omega_0^2 - \nu_m^2}{\omega_0^2 + \nu_m^2} \right| \approx \left| \frac{k_B T_e}{\nu_m} \frac{\partial \nu_m}{\partial \varepsilon} \right| << 1. \tag{48}$$

The numerical variation of the frequency dependent term is very weak and it can be approximated by unity. Considering finally $\nu_m = N_g \, v \, \sigma_m$ the condition reads:

$$\left| \frac{k_B T_e}{2\varepsilon} \left( 1 + \frac{2\varepsilon}{\sigma_m} \frac{\partial \sigma_m}{\partial \varepsilon} \right) \right| << 1. \tag{49}$$

For most gasses the cross section changes typically only slowly with energy and the derivative even vanishes at the Ramsauer minimum (typ. 0.5 eV) and at the maximum (typ. 10 eV), especially for the heavier noble gases. Other gases like helium, neon or hydrogen have a very flat cross section at energies below a few eV [33]. In these cases the condition reduces to $\left| k_B T_e / 2 \right| << \varepsilon$. On the other hand, while Eq. (49) indicates that the relative error is diverging for $\varepsilon \to 0$, the absolute error remains small since there the absolute collisionality is converging to zero and therefore does not play a crucial role. Last not least, the main intention of introducing the Fokker-Planck operator is the description of systems of low collisionality and not describing the highly collisional Ohmic and local limit. Therefore, one can conclude that the effect made by the finite velocity dependence of the collision frequency is small and the role of the function $g_1$ in the Fokker-Planck operator is probably negligible. However, this conclusion still needs be tested on real cross sections by using the Fokker-Planck operator in combination with a local Boltzmann solver. This will ultimately allow determining the effect on the non-local distribution function.

### IV. SURFACE LOSS OPERATOR

The remaining task is in the formulation of the surface loss operator. In a global model of a confined plasma, especially at low pressures where mean free paths are long, particle loss to the surface cannot be ignored. In particular, this particle loss to the surface has to balance creation of free charges in the volume by ionization. The loss rate depends on the plasma potential $\Phi$, which establishes itself in front of the surface and to a smaller extend within the volume. Only electrons with kinetic energies, in the coordinate perpendicular to the surface, higher than the potential can leave the plasma. The volume average of the loss across the total wall surface $A_w$ then reads:

$$
\begin{aligned}
\left.\frac{\partial f_0}{\partial t}\right|_{sur} &= -\left\langle \frac{A_w}{V} v \cos(\vartheta) f_0(v) \Theta\big(v\cos(\vartheta) - v_\Phi\big) \right\rangle_\Omega \\
&= -\frac{1}{2L'} \int_{v_\Phi/v}^{1} v\,\xi\, f_0(v)\, d\xi\, \Theta\big(v - v_\Phi\big) \\
&= -\frac{v}{4L'}\left(1 - \left(\frac{v_\Phi}{v}\right)^2\right) \Theta\big(v - v_\Phi\big) f_0(v).
\end{aligned}
\tag{50}
$$

The relevant length scale is $L' = V / A_w$. Only particles flowing towards the wall are lost, which restricts $1 \ge \cos(\vartheta) = \xi \ge 0$ in a local coordinate system with the $z$-axis pointing towards the surface. Only particles overcoming this potential can be lost, i.e. $v\xi \ge v_\Phi = \sqrt{2e|\Phi|/m}$. The factor $1/2$ results after $\varphi$ integration $(2\pi)$ and division by the entire solid angle of $4\pi$. In fact, the derivation is similar to the calculation of the current flowing to electrostatic probes [23], [33]. The operator has a form similar to the loss term for collisions with a natural threshold like excitation or ionization, although it approaches a constant for high energies and an effective cross section $\sigma_{loss}$ can be defined. The normalized effective cross section can also be interpreted as the ratio of the length $L'$ to the effective mean free path for surface loss $\lambda_{sur}$:

$$N_g \, \sigma_{loss} \, L' = \frac{L'}{\lambda_{sur}} = \frac{1}{4}\left(1 - \left|\frac{e\,\Phi}{\varepsilon}\right|\right) \, \Theta\left(\varepsilon - |e\,\Phi|\right). \tag{51}$$

Here $\varepsilon = m v^2 / 2$ is the kinetic energy. The plasma potential has to be found iteratively so that particle balance between ionization in the volume and loss to the surface is established ( $\sigma_{iz}$ is the ionization cross section and $N_g$ the neutral gas density):

$$\int_0^\infty \left(\frac{\partial f_0}{\partial t}\bigg|_{sur} + N_g \, \sigma_{iz}(v)\, v\, f_0\right) v^2 \, dv \overset{!}{=} 0. \tag{52}$$

The derivation of the above operator does not separate the plasma potential into the floating potential directly in front of the wall and the potential drop over the quasi-neutral plasma bulk. The potential drop over the bulk is of the order $k_B T_e$ and generally much smaller than the floating potential, typically by a factor 4 to 5. The actual value of the plasma potential found here will range somewhere between the value of the pure floating potential and the sum of the floating potential plus the smaller bulk potential. A more detailed modelling might be possible with some extra effort but is unlikely to lead to significant differences compared to the simplified approach proposed here.

## V. CONCLUSIONS AND OUTLOOK

Local collisional heating and non-local stochastic heating have been combined consistently in one Fokker-Planck operator. The two quite different electron-field interactions appear naturally as limiting cases in a systematic description. The Fourier transform approach allows a general formulation of the operator for arbitrary fields. Only an integral over the spatial Fourier transform of the electric field remains in the final expression for the coupling function. Depending on the particular spatial electric field profile, this integral might not always have analytical solutions. However, Boltzmann solvers integrate the Boltzmann equation numerically in any case so that a numerical table of the integral would probably be as good as an analytical formula. The resulting operator has a form similar to the heating operator from the classical two-term approximation. For a velocity independent collision frequency the form of the two operators is exactly identical but the local Ohmic coupling function $g_O$ is replaced by the generalized coupling function $g_2$ allowing also for non-local collisonless effects.

It has been suggested by Luis L. Alves that this similarity between the operators allows the definition of an effective cross-section for non-local heating [44]. The idea is to invert Eq. (4) in order to express $\nu_m = N_g v \sigma_m (v)$ as a function $g_O$. For the analogy now the Ohmic coupling function is replaced by $g_2$ in the collisonless limit $\nu_m \to 0$. This defines an effective cross section $\sigma_{eff}$ for the stochastic case:

$$N_g\,\sigma_{eff}\,\ell = \frac{\ell}{\lambda_{eff}} = \frac{v_\ell}{v} = \frac{1 - \sqrt{1 - \left(2g_2\left(\dfrac{v}{v_\ell}\right)\right)^2}}{2g_2\left(\dfrac{v}{v_\ell}\right)}. \tag{53}$$

As discussed already in section 3.2, $\ell$ is the characteristic gradient length responsible for the non-local effect and $v_\ell = \ell\,\omega_0$ is the corresponding characteristic velocity. Alternatively the expression for the effective cross section can be viewed as the inverse mean free path $\lambda_{eff}$ normalized by the gradient length $\ell$. For the particular case of the ICP discussed above $\ell = s$ (skin depth). Further, the coupling function (Eq. (45)) simplifies significantly for $\nu_m \to 0$. Here $\alpha = v/v_s$ is chosen in order to enhance the compactness of the presentation:

$$g_2(\alpha) = \frac{3}{\alpha}\left(\left(\frac{1}{2} + \frac{1}{\alpha^2}\right)\ln\left(1 + \alpha^2\right) - 1\right). \tag{54}$$

The example is shown in Fig.5. The single maximum and vanishing values at zero and infinite velocity contribute to a shape quite similar to classical gas-kinetic cross sections. For instance, for a gas density of $N_g = 10^{14}\,cm^{-3}$ and a skin depth of $s = 2\,cm$ the maximum effective cross section is $\sigma_{eff} = 9\cdot10^{-16}\,cm^2$, which is indeed of the same order as a classical gas-kinetic collision cross section. In fact, this is just an alternative way of highlighting the comparability in efficiency between Ohmic and stochastic heating.

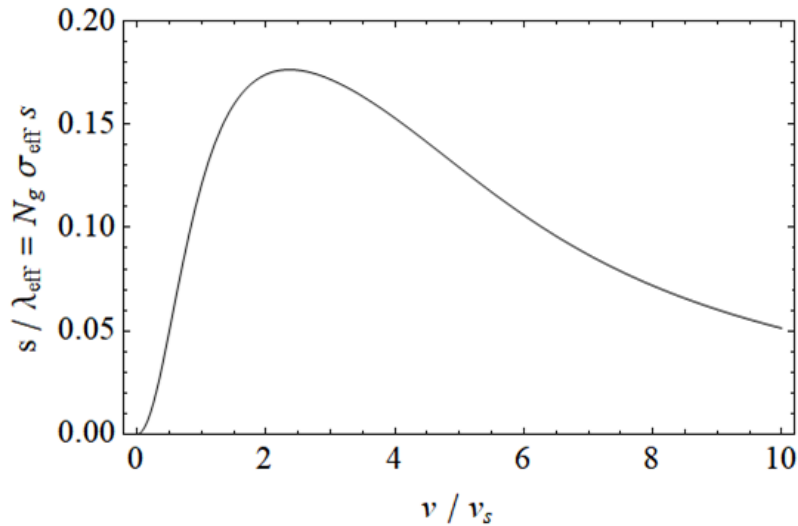

FIG. 5. Effective cross section for stochastic non-local heating as defined by Eq. (53) for the case of the classical ICP with skin depth $s$ and with $v_s = s\,\omega_0$.

In reality the collision frequency for momentum exchange is not constant but depends on velocity (or energy) in a way characteristic for each kind of gas. This leads to a second term in the Fokker-Planck operator related to the coupling function $g_1$. The possible impact of this additional term related to the velocity dependence of the collision frequency is analyzed by a criterion which defines conditions when the additional term can be neglected. Based on this criterion, one can conclude that the effect is probably negligible for most, if not all, gases.

The main part of the derivation is carried out for a single harmonic frequency. However, it is shown that the Fokker-Planck operator is linear in the different frequency components. For a field consisting of series of discrete frequencies a corresponding series of Fokker-Planck operators results. It is tempting to extend the present concept to continuous spectra and dispersion relations connecting $k$ and $\omega$. This is probably of less importance for the interaction with external fields, but would allow generalization to instabilities. Naturally, the dispersion relation and the related power spectrum of the instability need to be provided by a separate theory. In this case, the electron-field interaction does not necessary lead to heating but more likely thermal energy is converted to field energy which causes cooling. Further, the full extent of this option is only given by including also a static magnetic field. Certainly this is a very far reaching outlook.

The Fokker-Planck heating operator is an essential part of a larger concept to calculate the global non-local distribution function in low pressure plasmas by the use of a standard local Boltzmann solver. In the Boltzmann solver the classical heating operator resulting from the two term approximation is replaced by the Fokker-Planck operator. This replacement is straight forward since both operators have generally the same form but differ in the coupling functions. Further, the classical collisional loss operators have to be supplemented by a surface loss operator which accounts for particle and associated energy loss to the walls confining the plasma. This operator has also been derived and its form is similar to typical collisional loss operators describing threshold processes like excitation or ionization. Therefore, implementation of this operator should also be straight forward. The main new aspect related to the surface loss operator is in the self-consistent determination of the plasma potential. The potential needs to be adjusted iteratively in order to balance particle losses to the surface with particle generation by ionization. However, a sufficiently precise determination of the potential should be possible with the number of iterations not exceeding significantly about 10. Since modern local Boltzmann solvers are fast algorithms this additional iteration process should still keep the total calculation time at the order of seconds.

The main limitation of this concept is certainly in the requirement of know spatial profiles of the electric field. On the other hand the expected computational time is so fast that field structures for different scaling parameters or for different physical models can be quickly compared and the sensitivity of the distribution function on the detailed structure can be investigated. Further, the global distribution function is in many experimental cases a quantity that is well accessible to measurements, e.g. by Langmuir probes or Thomson scattering. Comparison will then be a crucial test of the models for the spatial field distribution. Last not least, comparison with more elaborate PIC/MC simulations is certainly necessary for testing the validity of the entire concept.

In the present formulation of the Fokker-Planck operator the power $h$-function is derived for arbitrary directions of the electric field. However, the final step of solid angle averaging, acting on the second order derivatives and the conductivity kernel, is carried out only for the case $\vec{E} \perp \vec{k}$. Arguably this is indeed the more versatile case but it does not include important applications like CCPs yet. Therefore, an important next step is the extension of the present concept also to fields which are spatially varying in the direction of the force. Further, so far the general Fokker-Planck operator has been derived without the presence of a static magnetic field $\vec{B}_0$. Extension of the above calculations by the Lorentz force should be straight forward, e.g. for ECR plasmas where $\vec{B}_0 \perp \vec{E}$.

In principle an equation similar to the Fokker-Planck equation should also follow from the Boltzmann equation by applying the two-term formalism also to the non-local case. The corresponding set of equations for $f_0$ and $f_1$ is sometimes called Davidov-Allis system. Reduction to a single equation for $f_0$ is often possible. Kolobov is carrying out this reduction and calls the resulting equation a Fokker-Planck equation [4]. However, no clear correlation with the present case is immediately visible which might be related to the fact that the equation is neither Fourier transformed nor volume averaged. Re-inspecting alternatively in Appendix A the set of two coupled equations following from the frequency analysis of the Boltzmann equation (Eq. (A3) and (A4)) it is rather straight forward to extend them in this sense. After Fourier transformation in space on the left hand side of Eq. (A4) an additional term $\vec{k} \cdot \vec{v}$ appears. Solving the equation for $f^{(1)}$ yields then the a term identical to the Fourier transform of the kinetic conductivity times the Fourier transform of the electric field and $f^{(0)}$. However, combining this intermediate result with Eq. (A3) is less straight forward. Fourier transformation of this equation leads to a convolution and further only the real part of above intermediate result enters here. Nevertheless, this is a direction worth to investigate more closely in future approaches.

In conclusion, a proposal has been made that should allow fast and direct calculation of the global non-local distribution function in low-pressure plasmas. Essential parts of the concept are a new Fokker-Planck heating operator, combining local Ohmic and non-local stochastic heating and a surface loss operator. Both operators have been derived under general conditions. The essential next step is the numerical implementation of the operators in an existing local Boltzmann solver, and comparison of the simulation results with alternative methods and/or experiments. This step has, at least in parts, been taken by the N-PRiME group of Lisbon, which has successfully implemented the proposed concept in the LisbOn KInetics Boltzmann solver (LoKI-B) [16] , [45].

**ACKNOWLEDGEMENT**

The author is indebted to Tsanko V. Tsankov, Luis L. Alves, and Wolfgang Paul for many inspiring discussions on this topic.

## APPENDIX A: LOCAL TWO-TERM APPROXIMATION FOR
## AN OSCILLATING FIELD

The ansatz for the distribution function in the Boltzmann Eq. (1) for a harmonic electric field $\vec{E} = E_0 \cos(\omega_0 t) \vec{e}_z$ is:

$$f(\vec{v},t) = \sum_{j=-\infty}^{\infty} f^{(j)}(\vec{v}) e^{ij\omega_0 t}, \tag{A1}$$

with $f^{(-j)} = f^{(j)*}$ in order to ensure the distribution function to be real. Then the Boltzmann equation turns into an iteration formula between the frequency components:

$$\left( i\,j - \frac{1}{\omega_0} \frac{\partial}{\partial t}\bigg|_{col} \right) f^{(j)} = \frac{v_E}{2} \frac{\partial}{\partial v_z} \left( f^{(j-1)} + f^{(j+1)} \right) \approx \frac{v_E}{2} \frac{\partial}{\partial v_z} f^{(j-1)}. \tag{A2}$$

Note that the operator $\partial/\partial t|_{col}$ is only an acronym for the Boltzmann collision integral where in addition summation is assumed to be carried out over all collision processes occurring in the plasma, i.e. elastic scattering, excitation, ionization etc.. Provided that $\alpha = v_E/v_{th} \ll 1$ , where $v_E = eE_0/(m\omega_0)$ is the velocity amplitude of a free oscillating electron and $v_{th}$ is the characteristic thermal velocity of the distribution, e.g. following from the mean energy, the various frequency components scale like $\left| f^{(j+1)}/f^{(j)} \right| = O(\alpha) \ll 1$ so that the series quickly converges. Then for $j \neq 0$ on the rhs $f^{(j+1)}$ can be neglected in good approximation and a convenient linear relation between $f^{(j)}$ and $f^{(j-1)}$ is established. The static case $j = 0$ is different. Noting that $f^{(-1)} + f^{(+1)} = 2\operatorname{Re}(f^{(1)})$ the corresponding equation reads:

$$-\frac{1}{\omega_0} \frac{\partial f^{(0)}}{\partial t}\bigg|_{col} = v_E \frac{\partial}{\partial v_z} \operatorname{Re}(f^{(1)}). \tag{A3}$$

According to Eq. (A2) the function $f^{(1)}$ is defined by:

$$\left( i - \frac{1}{\omega_0} \frac{\partial}{\partial t}\bigg|_{col} \right) f^{(1)} \approx \frac{v_E}{2} \frac{\partial}{\partial v_z} f^{(0)}. \tag{A4}$$

This is a closed set of coupled differential equations between the static and the first harmonic distribution function. Each frequency component can be expanded in a series of Legendre polynomials $P_k(\cos(\vartheta))$, where $\cos(\vartheta) = v_z/v$ and $\vartheta$ is the angle to the axis defined by the direction of the electric field. The particular structure of the derivative $\partial/\partial v_z = \cos(\vartheta)\,\partial/\partial v$ involves that $f_0^{(0)}$, the isotropic part of the time independent distribution, is only coupled to $f_1^{(1)}$, the anisotropic part of the distribution oscillating at $\omega_0$.

One can conclude further, that $f_k^{(0)}$ is non-zero only for even values of $k$ and $f_k^{(1)}$ only for odd values. Considering only the dominant contributions $k = 0$ and $k = 1$, respectively, the collision operator in Eq. (A4) simply yields the collision frequency for momentum loss $\nu_m$ summed over all collision types, i.e. dominantly elastic collisions but also to a minor extend momentum loss by inelastic collisions. This is the major step of the two-term approximation in this particular form of derivation:

$$\left. \frac{\partial}{\partial t} \right|_{col} f^{(1)} \approx -\nu_m f^{(1)}. \tag{A5}$$

Combining now Eq. (A4) with Eq. (A3) gives the final expression with $g_O$ defined as above in Eq. (4):

$$\frac{v_E^2}{2} \frac{\partial}{\partial v_z} \left( g_O \left( \frac{\nu_m}{\omega_0} \right) \frac{\partial}{\partial v_z} f_0^{(0)} \right) + \frac{1}{\omega_0} \left. \frac{\partial f_0^{(0)}}{\partial t} \right|_{col} = 0. \tag{A6}$$

Note that the collision integrals in the above equation, represented by the collision operator, are evaluated only for the isotropic part of the static distribution function. The equation can now be averaged over the solid angle in order to derive an equation for the isotropic part of the distribution $f_0^{(0)}$. For simplicity the upper index is dropped throughout this work and the understanding is that only the time independent part of the isotropic distribution function is discussed. Noting the particular form of the first derivative mentioned above, the second derivative becomes $\partial^2 / \partial v_z^2 = \cos^2(\vartheta) \partial^2 / \partial v^2 + \sin^2(\vartheta) / v \, \partial / \partial v$. Further, $\left\langle \sin^2(\vartheta) \right\rangle_\Omega = \left\langle \cos^2(\vartheta) \right\rangle_\Omega = 1/3$. The result of the solid angle average is then identical to Eq. (3). It should not pass unnoted that the result still holds for the static case, i.e. allowing $\omega_0 \to 0$, although it was derived for the oscillatory case.

## APPENDIX B: SOLID ANGLE AVERAGING OF THE FOKKER-PLANCK OPERATOR

In order to calculate the second order differential in the Fokker-Planck operator for the case $\vec{E} \perp \vec{k}$, it is assumed without loss in generality that the electric field points in $x$-direction and is spatially varying in $z$-direction. Eq. (8) then reads:

$$\left. \frac{\partial f_0}{\partial t} \right|_{FFPO} = \left\langle \frac{\partial^2}{\partial v_x^2} \left( h\left( v_z, \nu_m(v) \right) f_0(v) \right) \right\rangle_\Omega. \tag{B1}$$

Note that $\nu_m(v) = N_g v \sigma_m(v)$ is generally a function of the absolute velocity ($N_g$ is the neutral gas density). The second order differential with respect $v_x$ to has to be expressed in spherical coordinates $v, \xi = \cos(\vartheta), \varphi$ in order to carry out the solid angle average. The first order derivative is:

$$\frac{\partial}{\partial v_x} = \frac{\partial v}{\partial v_x}\frac{\partial}{\partial v} + \frac{\partial \xi}{\partial v_x}\frac{\partial}{\partial \xi} + \frac{\partial \varphi}{\partial v_x}\frac{\partial}{\partial \varphi}. \tag{B2}$$

The partial derivatives can be obtained noting that:

$$v_x = v\sqrt{1-\xi^2}\cos(\varphi), v_y = v\sqrt{1-\xi^2}\sin(\varphi), v_z = v\xi,$$

$$\Rightarrow v = \sqrt{v_x^2 + v_y^2 + v_z^2}, \ \xi = \frac{v_z}{v}, \ \varphi = \arctan\left(\frac{v_y}{v_x}\right). \tag{B3}$$

Then the first derivative reads:

$$\frac{\partial}{\partial v_x} = \sqrt{1-\xi^2}\cos(\varphi)\frac{\partial}{\partial v} - \xi\sqrt{1-\xi^2}\cos(\varphi)\frac{1}{v}\frac{\partial}{\partial \xi} - \frac{\sin(\varphi)}{\sqrt{1-\xi^2}}\frac{1}{v}\frac{\partial}{\partial \varphi}. \tag{B4}$$

Applying the same differential operator again yields in general 18 terms, which can be reduced to 9 different differentials. However, in the present case the function to be differentiated does not depend on $\varphi$, which drastically reduces the number of terms to 6 with only 5 different differentials. Further, the $\varphi$ average required for the final operator might be carried out straight away. Since all 5 terms contain only either $\sin^2(\varphi)$ or $\cos^2(\varphi)$ averaging over $\varphi$ yields a common factor $1/2$:

$$\left\langle\frac{\partial^2}{\partial v_x^2}\right\rangle_\varphi = \frac{1}{2v^2}\left(\xi\left(1-3\xi^2\right)\frac{\partial}{\partial \xi} + \xi^2\left(1-\xi^2\right)\frac{\partial^2}{\partial \xi^2}\right.$$

$$\left. -2\xi\left(1-\xi^2\right)v\frac{\partial^2}{\partial \xi \partial v} + \left(1+\xi^2\right)v\frac{\partial}{\partial v} + \left(1-\xi^2\right)v^2\frac{\partial^2}{\partial v^2}\right). \tag{B5}$$

The operator averaged over $\xi$ reads after removing the differentials by partial integration:

$$\left\langle\frac{\partial^2}{\partial v_x^2}\right\rangle_\Omega = \frac{1}{4v^2}\left(\int_{-1}^{1}\left(1-3\xi^2\right)h\,d\xi + v\frac{\partial}{\partial v}\int_{-1}^{1}\left(3-5\xi^2\right)h\,d\xi + v^2\frac{\partial^2}{\partial v^2}\int_{-1}^{1}\left(1-\xi^2\right)h\,d\xi\right). \tag{B6}$$

This has the general form of:

$$\left.\frac{\partial f}{\partial t}\right|_{FFPO} = \frac{1}{4v^2}\left(\Gamma_0 f_0 + v\frac{\partial}{\partial v}\left(\Gamma_1 f_0\right) + v^2\frac{\partial^2}{\partial v^2}\left(\Gamma_2 f_0\right)\right). \tag{B7}$$

The functions $\Gamma_{0,1,2}$ can be readily identified:

$$\Gamma_0(v) = \int_{-1}^{1} \left(1 - 3\xi^2\right) h(v\xi, \nu_m) d\xi,$$

$$\Gamma_1(v) = \int_{-1}^{1} \left(3 - 5\xi^2\right) h(v\xi, \nu_m) d\xi, \qquad \text{(B8)}$$

$$\Gamma_2(v) = \int_{-1}^{1} \left(1 - \xi^2\right) h(v\xi, \nu_m) d\xi.$$

Particle number conservation requires that integration over velocity $\left(v^2 dv\right)$ should yield zero. After partial integration the resulting relation is:

$$\Gamma_0 = \Gamma_1 - 2\Gamma_2. \qquad \text{(B9)}$$

Inserting the $\Gamma_j$ functions identified above fulfills indeed this requirement. Consequently, there are only two and not three independent functions. This motivates to express the Fokker-Planck operator in a form close to the operator following from the two-term approximation:

$$\frac{\partial f}{\partial t}\bigg|_{FFPO} = \frac{\omega_0 \left\langle v_E^2 \right\rangle_V}{6v^2} \frac{\partial}{\partial v} \left( g_1(v) v f_0 + g_2(v) v^2 \frac{\partial}{\partial v} f_0 \right). \qquad \text{(B10)}$$

Comparison of the terms yields:

$$g_1(v) = \frac{3}{2h_0} \left( \Gamma_1 - 2\Gamma_2 + v \frac{\partial}{\partial v} \Gamma_2(v) \right),$$

$$g_2(v) = \frac{3}{2h_0} \Gamma_2(v), \qquad \text{(B11)}$$

$$h_0 = \omega_0 \left\langle v_E^2 \right\rangle_V = \omega_0 \left\langle \left| \frac{e\vec{E}_0}{m\omega_0} \right|^2 \right\rangle_V.$$

Note that when using in the calculation of $h_0$ a field structure that is artificially extended to the negative half sphere in order to account for surface reflection of electrons, the volume $V$ which is used in the average also doubles, i.e. $V \rightarrow 2V$. Naturally, nothing changes if the original field structure existing only in the positive half-sphere is used and both averages are identical in any case.

In the velocity derivative appearing in $g_1$ use can be made of the particular dependence of $h = h(v\xi, \nu_m(v))$ on the variables:

$$v\frac{\partial}{\partial v}\Gamma_2(v) = \int_{-1}^{1}(1-\xi^2)v\frac{\partial}{\partial v}h(v\xi,\nu_m)d\xi$$

$$= \int_{-1}^{1}(1-\xi^2)v\left(\frac{\xi}{v}\frac{\partial}{\partial\xi}h(v\xi,v\sigma) + \frac{\partial\nu_m}{\partial v}\frac{\partial}{\partial\nu_m}h(v\xi,\nu_m)\right)d\xi$$

$$= \int_{-1}^{1}(1-\xi^2)\xi\frac{\partial}{\partial\xi}h(v\xi,v\sigma)d\xi + v\frac{\partial\nu_m}{\partial v}\frac{\partial}{\partial\nu_m}\Gamma_2 \qquad \text{(B12)}$$

$$= -\int_{-1}^{1}(1-3\xi^2)h(v\xi,v\sigma)d\xi + v\frac{\partial\nu_m}{\partial v}\frac{\partial}{\partial\nu_m}\Gamma_2$$

$$= -\left(\Gamma_1(v) - 2\Gamma_2(v)\right) + v\frac{\partial\nu_m}{\partial v}\frac{\partial}{\partial\nu_m}\Gamma_2.$$

This allows expressing $g_1(v)$ by the derivative of $g_2(v)$:

$$g_1(v) = v\frac{\partial\nu_m}{\partial v}\frac{\partial g_2}{\partial\nu_m} = \left(1 + \frac{v}{\sigma_m}\frac{\partial\sigma_m}{\partial v}\right)\nu_m\frac{\partial g_2}{\partial\nu_m},$$

$$g_2(v) = \frac{3}{2}\int_{-1}^{1}(1-\xi^2)\frac{h(v\xi,\nu_m)}{h_0}d\xi. \qquad \text{(B13)}$$

Depending on convenience, the preceding factor in $g_1$ might be expressed either by the derivative of the collision frequency $\nu_m = N_g v\sigma_m(v)$ or the collision cross section $\sigma_m(v)$ for momentum changing collisions. If $h$ is entirely Ohmic then indeed $g_2 \propto g_o$. If further $\nu_m$ is constant, apparently $g_1$ vanishes. In this case the structure of Eq. (B10) is identical to the two-term approximation result but $g_2$ replaces $g_o$. In conclusion, in the Ohmic case differences arise only by the velocity dependence of $\nu_m$ which leads to the appearance of an addition function $g_1$. In the main text, the concept is further extended by applying the derivatives and integrations directly to the conductivity kernel. The final expressions then contain only integrals over $k$.